\documentclass{WileyMSP-template}

\usepackage{amssymb,amsmath,amsthm,amsfonts,amscd}
\usepackage{graphicx}
\usepackage[dvipsnames]{xcolor}
\usepackage{mathrsfs}
\usepackage[utf8]{inputenc}
\usepackage{tikz-cd}
\usepackage[english]{babel}
\usepackage{dsfont}
\usepackage{url}
\usepackage{nicematrix}

\DeclareMathOperator{\tr}{tr}

\renewcommand{\Im}{\mathrm{Im}}

          \theoremstyle{definition}

\newcommand{\eq}[1]{\begin{equation}#1\end{equation}}

\newcommand{\ee}{\mathrm{e}}

\usepackage{comment}
\usepackage{graphicx}
\usepackage{braket}
\usepackage{color}

\definecolor{mygreen}{rgb}{0,0.5,0}
\definecolor{myblue}{rgb}{0,0,0.75}
\definecolor{mymagenta}{cmyk}{0,1,0,0.12}
\usepackage[unicode,psdextra,colorlinks=true,
   linkcolor=blue,
   citecolor=blue,
   urlcolor =blue]{hyperref}

\usepackage[sort&compress,capitalise,nameinlink]{cleveref}
\usepackage[longnamesfirst,numbers,sort&compress]{natbib}
\crefname{section}{\textsection}{\textsection}
\crefname{subsection}{\textsection}{\textsection}
\crefname{subsubsection}{\textsection}{\textsection}
\crefname{paragraph}{\textparagraph}{\textparagraph}
\crefname{thm}{Theorem}{Theorems}
\usepackage{amsmath}

\begin{document}
\tolerance=2000
\setlength{\emergencystretch}{1em}
\pagestyle{fancy}


\newcommand{\btext}[1]{{\color{myblue} #1}}
\newcommand{\gtext}[1]{{\color{mygreen} #1 }}
\newcommand{\mtext}[1]{{\color{mymagenta} #1}}
\newcommand{\citeSM}{\cite[{\tiny SM}\kern-0.3em][]{SM}}
\newcommand{\hc}{\widetilde{\mathrm{h.c.}}}
\newcommand{\be}{\begin{equation}}
\renewcommand{\ee}{\end{equation}}
\newcommand{\rtext}[1]{{\textcolor{red}{#1}}}
\newcommand{\cf}{{\it cf.}}
\newcommand{\ie}{{\it i.e.}}
\newcommand{\Prob}{\mathrm{Prob}}
\renewcommand{\tr}{\mathrm{tr}\,}
\newcommand{\bv}[1]{{\color{myblue} BV: #1}}

\expandafter\let\csname equation*\endcsname\relax

\expandafter\let\csname endequation*\endcsname\relax

\pagestyle{fancy}


\pagestyle{fancy}

\title{Entanglement Hamiltonians - from field theory, to lattice models and experiments}

\maketitle


\author{Marcello Dalmonte$\,^1$, Viktor Eisler$\,^2$, Marco Falconi$\,^3$, Beno\^it Vermersch$\,^{4,5,6}$}

\begin{affiliations}
$^1$The Abdus Salam International Centre for Theoretical Physics, Strada Costiera 11, 34151 Trieste, Italy, and SISSA, via Bonomea 265, 34136 Trieste, Italy. \href{mailto:mdalmont@ictp.it}{mdalmont@ictp.it}
\\[2mm]
$^2$Institute of Theoretical and Computational Physics, Graz University of Technology, Petersgasse 16, 8010 Graz, Austria. \href{mailto:viktor.eisler@tugraz.at}{viktor.eisler@tugraz.at}
\\[2mm]
$^3$Dipartimento di Matematica, Politecnico di Milano, Piazza Leonardo da Vinci 32, 20133 Milano, Italy. \href{mailto:marco.falconi@polimi.it}{marco.falconi@polimi.it}
\\[2mm]
$^4$Universit\'e  Grenoble Alpes, CNRS, LPMMC, 38000 Grenoble, France. 
\\
$^5$Center for Quantum Physics, University of Innsbruck, Innsbruck A-6020, Austria.
\\
$^6$Institute for Quantum Optics and Quantum Information of the Austrian Academy of Sciences,  Innsbruck A-6020, Austria.
\href{mailto:benoit.vermersch@lpmmc.cnrs.fr}{benoit.vermersch@lpmmc.cnrs.fr}
\\[3mm]


\end{affiliations}


\begin{abstract}

We review results about entanglement (or modular) Hamiltonians of quantum many-body systems in field theory and statistical mechanics models, as well as recent applications in the context of quantum information and quantum simulation. 

\end{abstract}

\section{Introduction}

Since its inception, quantum mechanics has routinely thrilled theoretical and experimental physicists due to its striking - and, at times, counter-intuitive - differences from the classical world one is accustomed to. One of the most remarkable examples of those is certainly entanglement~\cite{Nielsen2011,casatiBook}: this form of quantum correlations went relatively unnoticed until the last decades of the twentieth century, just to experience a burst of interest and applications since.

Entanglement captures the degree of inseparability between quantum mechanical states: it vanishes in the absence of quantum correlations, and it is instead finite in case separability is not satisfied. It has found numerous and profound applications in quantum information science, ranging from quantum communication to its connection to the capabilities of quantum computing hardware, just to name a few~\cite{Nielsen2011,casatiBook,Gardiner_vol3}. 

In parallel to progresses in quantum information and quantum optics, it has emerged that entanglement is not only useful to describe few-body processes at a microscopic level, but that, in addition, it is an invaluable resource in characterizing and classifying genuine quantum mechanical features in many-body systems~\cite{Amico2008,CCD09,Eisert2010,Laflorencie2016,moessner_moore_2021,witten2018arxiv}. Nowadays, entanglement is routinely used in a variety of contexts: since its first applications in many-body lattice models - in particular, focused on critical behavior~\cite{Amico2008,CCD09,Eisert2010} and topological order~\cite{Laflorencie2016,moessner_moore_2021} -, it has found applications as wide as the characterization of computational methods and their performances~\cite{Schollwock2011}, to the classification of many-body behavior and holographic properties in quantum field theory~\cite{Calabrese_2009,Casini_2009,Nishioka_2009,witten2018arxiv}. 

\begin{figure}[t]
\centering
\includegraphics[width=0.95\textwidth]{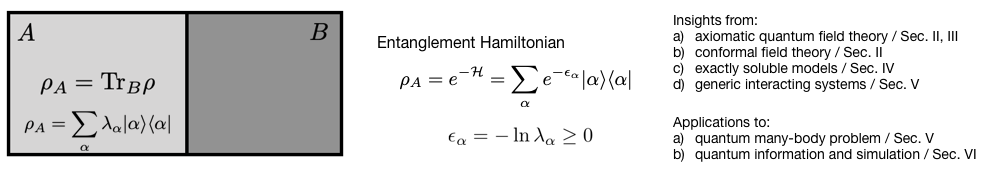}
\caption{Playground of this review: given a quantum mechanical state $\rho$, and two spatial partitions $A$ and $B$, we are interested in determining the properties of $\rho_A$ in many-body systems.}
\label{fig:cartoon}
\end{figure}

The main actors in determining the relevance of entanglement in many-body systems are measures of bipartite entanglement (and, to a lesser extent, entanglement witnesses and measures of multipartite entanglement)~\cite{casatiBook,PV07}. Consider as an example the setting described in Fig.\ref{fig:cartoon}: given a state $\rho$, one is interested in characterizing the quantum correlations between its partition $A$ and its complement $B$. Those are captured by the reduced density matrix (RDM):
\begin{equation}
    \rho_A = \text{Tr}_B\, \rho = \sum_{\alpha=1}^{r_A}\lambda_\alpha |\alpha\rangle \langle\alpha|
\end{equation}
where the last equality represents an eigendecomposition with eigenvalues $\lambda_\alpha$ and eigenvectors $|\alpha\rangle$, and $r_A$ is the rank of $\rho_A$~\footnote{This representation is directly related to Schmidt decomposition for pure states.}. 

For the case of globally pure states - that is, states for which $\text{Tr}\rho^2 = 1$ - a paramount quantity that quantifies the entanglement between $A$ and $B$ is the von Neumann entropy~\cite{Bennett1996}. The von Neumann entropy and other related measures (such as concurrence~\cite{Amico2008}) have found extensive use in quantum many-body theory - see, e.g., the reviews in Ref. ~\cite{Amico2008,Calabrese_2009,witten2018arxiv}. One key aspect of the entropy and related quantities is that they are solely sensitive to the eigenspectrum of reduced density matrix - the so-called entanglement spectrum~\cite{Li2008,Laflorencie2016}. The much richer relation between the finer structure of the RDM - that is, the structure of the eigenvectors, and their relation to the corresponding eigenvalues - has instead been initially overlooked, partly because it is considerably more challenging to characterize at the theoretical (and experimental) level. It is this finer structure that is the topic we plan to cover in this review.

Our starting point is the observation that, given the fact that a density matrix is always positive semi-definite, it is always possible to cast it in an exponential form (analogously to a `partition function') as follows:
\begin{equation}
    \rho_A \propto e^{-\mathcal{H}}.
\end{equation}
The dimensionless operator $\mathcal{H}$ is known under two different names: modular Hamiltonian in quantum field theory, and entanglement Hamiltonian (EH) in the context of lattice models and quantum information. Throughout this work, we will follow the second nomenclature. The operator $\mathcal{H}$ is bounded from below, as a consequence of the semi-positive-definiteness of the reduced density matrix.

The entanglement Hamiltonian has a somehow interesting history, that proceeds along three parallel routes. Historically, the first appearance of EH was related to developments in the field of axiomatic field theory (1970's): the most relevant achievement in that field was a set of results that are referred to as Bisognano-Wichmann theorem~\cite{bisognano1975jmp,bisognano1976jmp}, that presented a mathematically rigorous computation of the EH within certain - rather generic - field theoretical settings. This was shortly followed up by works dealing with conformal field theories~\cite{hislop1982cmp}. The second advent of the EH is related to integrable models (1980's): within certain of those settings, the EH has very strong ties with the corner transfer matrix~\cite{PKL99}, a pivotal object in the field of lattice statistical mechanics~\cite{baxterbook}. The third one is related to topological matter ((2000's))~\cite{moessner_moore_2021}: there, the spectrum of the EH was identified as a very potent tool to diagnose and characterize topological phases~\cite{Li2008}, both at the level of true topological (e.g., fractional quantum Hall effect) and of symmetry-protected topological order (e.g., topological insulators).

Interestingly, the three aforementioned routes proceeded almost independently, with very few intersections, until very recently. In particular, developments in the field of quantum field theory have been almost completely decoupled from lattice statistical mechanics models.

Against this background, the aim of this review is two-fold. First, we will present the basic toolkit and most representative results from those three separate research lines, and emphasize the net of connections between all of those that has been developed over the last years. The quantum field theory side of the story will be presented in Secs. ~\ref{Sec:Defs} and ~\ref{sec:aQFT}: we will start from a brief collection of all relevant results in the field, and then proceed with a more mathematical oriented derivation of the main tool available - Tomita-Takesaki theory - at the basis of the Bisognano-Wichmann theorem. In Sec.~\ref{sec:int}, we will discuss integrable lattice models: these models allow for considerable analytical insight into the structure of EH, that is of key importance to gauge the relevance of field theory results on the lattice. This discussion will naturally evolve to non-integrable lattice models in Sec.~\ref{sec:lattice}. There, we will emphasize the role played by entanglement Hamiltonians in the context of topological and quantum critical matter. 
The second aim of this review is to present applications of such knowledge in the broader context of quantum information science. Sec.~\ref{sec:Exp} will be devoted to this task: in particular, we will review methods to characterize many-body systems in quantum computers and quantum simulations that leverage on the many-body insights based on entanglement Hamiltonians. 

While, as mentioned above, different communities have studied entanglement Hamiltonians with limited crosstalks, we have opted to follow a different route. While we have kept the various lines separate in terms of sections, throughout the review, we have often commented on how results in one field have impacted others, both in terms of interpretation of the physical results, and methods. Hopefully, such choice may help the reader interested in a specific section to better appreciate the breadth of the field, and the various intersections.

The entanglement Hamiltonian is of course not disconnected from entropies and the entanglement spectrum: the first are nothing but moments of its expectation value, and the latter is its spectrum. Since these quantities have already been subjects of excellent reviews (see Ref.~\cite{Calabrese_2009,Eisert2010,Laflorencie2016}), the stress of our article will be on aspects of the EH that go beyond those - in particular, its structure as a many-body operator. While not extensively, we will refer to properties of the ES whenever useful.

\section[Insights from quantum field theory]{Entanglement Hamiltonian: definitions and insights from quantum field theory}\label{Sec:Defs}

In this section, we will first provide some definitions and basic observations on entanglement Hamiltonians and related quantities. Then, we will review some basic results about EH of quantum field theories. Some of those will then be reframed in a more rigorous mathematical framework in the next section.

\subsection{Definitions and notations}

Let us recall here some basic properties of the reduced density matrices we are dealing with. First of all, we will always work with normalized RDMs, 
\begin{equation}
    \text{Tr}\,\rho_A = 1, \quad \sum_\alpha \lambda_\alpha=1.
\end{equation}
This will require some care in defining the EH: indeed, shifting the operator $\mathcal{H}$ by a constant will change the trace of $\rho_A$. We will utilize the following normalization:
\begin{equation}
        \rho_A = e^{-\mathcal{H}} / Z 
\end{equation}
which is related to another one used in literature as follows:
\begin{equation}
        \rho_A = e^{-(\mathcal{H} + C)}, \quad  Z = \exp{C},
\end{equation}
where $C$ is a constant. This consideration implies that the absolute values of the entanglement spectrum $\epsilon_\alpha$ have little operational meaning, while what is really relevant are the corresponding entanglement energy differences and relative magnitudes. This fact will have direct implications in protocols to measure the ES experimentally. 

It is worth noting that, at first sight, it is very unclear whether the EH shall have a meaningful many-body structure. Indeed, one could simply interpret it as the logarithm of a density matrix. In general, the corresponding operator does not have to be local, neither few-body\footnote{A notable exception here are Gaussian states, where, due to Wick's theorem, the EH must only contain two-body terms. This will be at the core of the insights presented in Sec~\ref{sec:int}}.

It is thus a remarkable fact that there are {\it generic} circumstances under which the EH of an extended partition $A$ of a many-body system is extremely well approximated - if not exactly given - in terms of operators $\mathcal{H}(x)$ that are both local and few-body:
\begin{equation}
    \mathcal{H} = \int_{x\in A} dx \, \beta(x) \, \mathcal{H}(x) \, ,
    \label{eq:EHbeta}
\end{equation}
where $\beta(x)$ is some weight function.
The reason behind this fact traces back to quantum field theory (that naturally encompasses locality, relativistic invariance, and quantum mechanical effects), and is centered around the locality of many-body dynamics. 

\subsection[Bisognano-Wichmann theorem]{Entanglement Hamiltonian and the Bisognano-Wichmann theorem}

The most basic result at the heart of the few-body, local nature of EH is the Bisognano-Wichmann theorem~\cite{bisognano1975jmp,bisognano1976jmp}. Consider a relativistic quantum field theory in (D+1)-dimensions ($D$ denotes the spatial dimensions only), with Hamiltonian density $H(x)$, and with coordinates $x = \{x_1, x_2, .., x_D\}$ labeling  spatial coordinates. Given a half-plane partition $A$ defined by the condition $x_1>0$, the Bisognano-Wichmann theorem states that the entanglement Hamiltonian of the vacuum state reads:
\begin{equation}
    \mathcal{H} = \frac{2\pi}{c}\int_{x\in A} dx \; x_1 H(x)\label{eq:BW}
\end{equation}
where $c$ is the corresponding `speed of light', which makes the EH dimensionless. This result - applicable irrespective of the particle content of the theory, and compatible with the existence of gauge symmetries - establishes that not only is the EH of such states local and few-body: it is actually nothing but the boost operator, that is, the original Hamiltonian with space dependent couplings. 

\begin{figure}[t]
\centering
\includegraphics[width=0.95\textwidth]{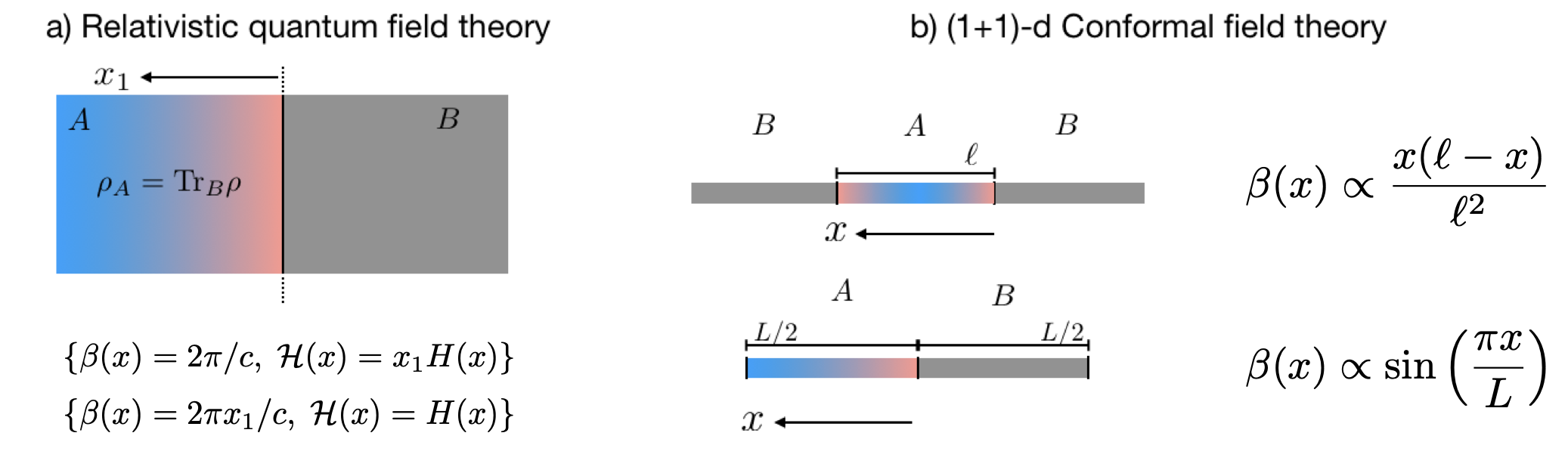}
\caption{Bisognano-Wichmann theorem and entanglement Hamiltonian in quantum field theory. Panel {\it a)}: for a generic, relativistic quantum field theory, the entanglement Hamiltonian of a half-system is given by the boost operator. The corresponding reduced density matrix can be interpreted as either a thermal state with Hamiltonian density $x_1H(x)$, where $x_1$ is the distance from the boundary, or with an equilibrium state subject to a position dependent inverse temperature $\beta(x) =  2\pi x_1/c$, and Hamiltonian density $H(x)$. In the second interpretation, the entanglement temperature goes from 'high' close to the boundary between the partitions (red), to cold away from the boundary (blue). Panel {\it b)}: some of the extensions of the Bisognano-Wichmann to 1+1 dimensional conformal field theory, formulated in terms of a space-dependent inverse entanglement temperature: finite interval in an infinite system (top), and half-chain with open boundary conditions (bottom).}
\label{fig:cartoonBW}
\end{figure}

A first physical interpretation of the Bisognano-Wichmann theorem is illustrated in Fig.~\ref{fig:cartoonBW}a: one can either see Eq.~\eqref{eq:BW} as describing a thermal state with respect to the boost operator, or, equivalently, describing a thermal state with respect to the original Hamiltonian, but with a position dependent temperature. In this second interpretation, the entanglement temperature is `high' close to the boundary, and progressively decreases moving away from it. This interpretation immediately suggests that the quantum correlations between $A$ and its complement are overwhelmingly dominated by the `high' region - the boundary -, thus providing a simple picture to understand a conjecture (proved in $D=1$ for the case of gapped theories \cite{Hastings07}) known as area law \cite{Eisert2010}. At the end of the next section, we will discuss another physical interpretation in the context of the Unruh effect~\cite{Unruh76} (somehow related to the first viewpoint above).

\subsection[Conformal symmetry]{Entanglement Hamiltonians in the presence of conformal symmetry}\label{sec:CFTs}

In the presence of additional global conformal symmetry, the result above can be extended to different systems, and different types of partitions. Here, we review the basic extensions of the BW theorem that have been obtained so far. 

\subsubsection{$D$-dimensional theories}

For a partition $A$ consisting of a ball of volume $V$ and radius $R$ in a $D$-dimensional relativistic conformal field theory, it was shown in Ref. \cite{hislop1982cmp,CHM11} that the corresponding entanglement Hamiltonian reads:
\begin{equation}
    \mathcal{H}_A^{CFT0} = \frac{2\pi R}{c} \int_{x\in A} dx\; \frac{R^2-r^2}{2R^2} T_{00}(x) \label{eq:CHM}
\end{equation}
where the ball has been placed at the origin, $r(x)$ is the Euclidean distance from the origin, $T_{00}$ is the stress-energy tensor of the CFT (in fact analogue of the Hamiltonian density). The physical consequence of this result is that, in the presence of conformal symmetry, the EH retains all its useful locality and few-body properties even in cases of finite partitions, even if it is not correspondent to the boost operator any more. 
Note that the treatment of more complicated partitions (other than the half-space and the sphere) in $D>1$ is very complicated in general, however, some approximate approaches exist for free field theories, predicting the emergence of anisotropic inverse entanglement temperatures \cite{ACHP17}. 

\subsubsection{One-dimensional theories}

In $D=1$ spatial dimension, conformal symmetry becomes even more powerful in allowing the determination of EH for various types of partitions. This was originally explored in \cite{hislop1982cmp}, and further extended in Ref.~\cite{CardyTonni16}. We review here the basic cases of interest to equilibrium physics, noting that, remarkably, some extensions to time-dependent problems have also been derived ~\cite{CardyTonni16,Wen_2018}.

The first relevant case deals with a partition of finite size $\ell$, embedded in an infinite chain (Fig.~\ref{fig:cartoonBW}b). In that case, the EH reads:
\begin{equation}
    \mathcal{H}_{A}^{CFT1} = \frac{2\pi \ell}{c}\int_{0}^\ell dx \; \left[\frac{x(\ell-x)}{\ell^2} H(x) \right] 
    \label{eq:CFT1}
\end{equation}
that corresponds to the (1+1)-d version of Eq.~\eqref{eq:CHM}. Note that, close to each of the boundaries, the EH is reminiscent of the original BW functional form. By exploiting conformal mappings, it is possible to generalize this formula to the case of a finite interval inside a finite circle of circumference $L$, that then reads:
\begin{equation}
    \mathcal{H}_{A}^{CFT2} = \frac{2L}{c}\int_{0}^\ell dx \; \left[ \frac{\sin\left(\frac{\pi (\ell-x)}{L} \right) \sin\left(\frac{\pi x}{L} \right)}{\sin\left(\frac{\pi \ell}{L} \right)} H(x) \right].
    \label{eq:CFT2}
\end{equation}
This last equation will be particularly useful in the numerical studies
in Sec. \ref{sec:lattice}. 

Lastly, it is possible to consider a finite interval of length $L/2$ at the boundary of an open chain of length $L$ (Fig.~\ref{fig:cartoonBW}b): in this case, the EH reads:
\begin{equation}
    \mathcal{H}_{A}^{CFT3} = \frac{2L}{c}\int_{0}^{L/2} dx \;  \sin \left(\frac{\pi x}{L}\right)H(x).  
    \label{eq:CFT3}
\end{equation}
The physical interpretation of these results is immediate, and graphically illustrated in Fig.~\ref{fig:cartoonBW}: all of them predict $\beta(x)$ to be linear close to the boundaries, i.e. the entanglement temperature decreases with the inverse of the distance within the 'entangled' region, whereas $\beta(x)$ saturates following a 'parabolic-like' shape at large distances, consistently with the generic prediction in Ref. \cite{CHM11}.

\subsection{Entanglement Hamiltonian at finite temperature}

So far, we have only dealt with vacuum states of a field theory - that is, pure states. Remarkably, under some specific conditions, it is also possible to get insights on the structure of the EH for the case of mixed states. 

The first example are thermal states at inverse temperature $\beta$. If one considers a small partition embedded in a large system whose dynamics is described by a non-integrable theory, the corresponding entanglement Hamiltonian is determined by quantum typicality, and it is nothing but the Hamiltonian of the subsystem $H_A$, that is, $\rho_A\propto e^{- \beta H_A}$, up to corrections due to boundary terms. This formula is expected to work well for temperatures larger than the typical energy scales in the system. 

Remarkably, again for the case of one-dimensional conformal field theories, even more precise statements can be made at {\it any} temperature~\cite{borchers1999,Wong2013}. In that case, the EH of a finite region embedded in an infinite system reads:
\begin{equation}
    \mathcal{H}_{A}^{CFT4} = 2\beta\int_{0}^\ell dx \; \left[ \frac{\sinh\left(\frac{\pi (\ell-x)}{\beta c} \right) \sinh\left(\frac{\pi x}{\beta c} \right)}{\sinh\left(\frac{\pi \ell}{\beta c} \right)} H(x) \right].
    \label{eq:CFT4}
\end{equation}
The result above reproduces Eq. \eqref{eq:CFT1} in the limit $\beta\gg \ell$, where the EH properties are well captured by the BW theorem. On the other hand, in the limit $\beta \ll \ell$ one approaches a Gibbs state at inverse temperature $\beta$. This prediction would also be obtained, assuming typicality, for pure states of the total system at high energies. We note that some of these results can also be extended to the post-measurement case, as has been show in Ref.~\cite{Rajabpour2016}.

\subsection{Further results\label{sec:CFTfurther}}

The examples discussed so far are valid for generic CFTs. Further results exist for some specific field theories, such as the free boson or fermion theories. In particular, the case of multiple disjoint intervals has been derived for the massless Dirac fermion \cite{CasiniHuerta09} and free scalar theories \cite{ACHP18}, where the local part of the EH has again a similar structure \eqref{eq:EHbeta} as in the single interval case. Namely, the energy density is multiplied with an appropriate space-dependent weight function $\beta(x)$ which can be obtained explicitly \cite{CasiniHuerta09,Longo10,ABCH17,ACHP18}. Some of these results can even be extended to finite temperatures \cite{Fries2019,Blanco19,Hollands21}. For the Dirac theory the main new feature is a subleading \emph{bi-local} piece in the EH, which couples only to a single conjugate point in each of the intervals. Similar features have been observed for the massless Dirac fermion in the presence of a boundary or defect \cite{MintchevTonni21a,MintchevTonni21b}. The explicit treatment of the massive case is much more difficult even for a single interval, and the available results suggest the presence of a subleading \emph{nonlocal} piece in the EH \cite{ABCH17,LongoMorsella20}. Remarkably, similar nonlocal terms were found to appear also for massless fermions in the presence of zero modes \cite{KVW17} and in low-lying excited states \cite{KVW18}.

Finally, one should mention that some extensions to \emph{inhomogeneous} physical systems
exist, the simplest example being the massless Dirac fermion with a spatially varying Fermi velocity \cite{Tonni18}. This inhomogeneity can be absorbed by a local change of the metric and is equivalent to considering the Dirac theory in a curved background \cite{DSVC17}. Considering the EH of a singly connected bipartition, the effect of the curved metric can be taken into account and leads again to a local expression \eqref{eq:EHbeta} where $\beta(x)$ can be obtained explicitly \cite{Tonni18}.

\vspace{5mm}
We have now provided a full picture about what is known about EH in quantum field theory. The pivotal result throughout our discussion has been the Bisognano-Wichmann theorem, both as a result per se, and as a starting point for extensions to conformal field theory. In the following section, we will address the question: why is the EH taking such a simple functional form? In particular, what is the principle that determines its locality? Those questions are properly answered in the context of axiomatic quantum field theory, that we will try to present in a concise form that is of use to theoretical physics. Readers that are not familiar with QFT can proceed directly to Sec. \ref{sec:int}, where we discuss lattice statistical mechanics models.

\section[Tomita-Takesaki and Bisognano-Wichmann]{Tomita-Takesaki's modular theory and Bisognano-Wichmann's theorem in
  algebraic QFT: mathematics comes to rescue}\label{sec:aQFT}

In this section we will briefly overview two interconnected mathematical
aspects of quantum theories that play a crucial, if somewhat unexpected, role
in the description of entanglement Hamiltonians: Tomita-Takesaki theory of
modular groups, and Bisognano-Wichmann's theorem in algebraic QFT. The
mathematical details will be kept to a minimum, the interested reader may
refer to several books and reviews on algebraic quantum field theory, and the
references thereof contained; among those, let us single out
\citep{bratteli1987tmp1, bratteli1997tmp2, haag1992tmp, borchers2000jmp},
from which we drew inspiration in writing this review, and the more
physically oriented review by Witten \citep{witten2018arxiv}.

\subsection{Tomita-Takesaki's theory}
\label{sec:tomita-takes-theory}

This theory has been developed by Tomita in the late sixties \citep[the
results were announced in][]{tomita1967sendai}, and then expanded and made
known to a broad audience by Takesaki with his 1970 lecture notes
\citep{takesaki1970lnm}.

The theory of Tomita and Takesaki provides a way to construct, given an
algebra of quantum observables $\mathcal{M}$ realized as operators on some
Hilbert space $\mathscr{H}$, and a quantum state realized as a vector
$\Omega$ (with some `reasonable' properties, to be detailed below), two
operators associated with the couple $(\mathcal{M},\Omega)$: the modular operator
$\Delta$, and the modular conjugation $J$. These operators enjoy some nice
properties, again to be detailed below.

The importance of modular operators $\Delta$ and $J$ in characterizing algebras of
operators, both as pure mathematical objects and as collections of quantum
observables, has very soon become apparent. In fact, on one hand
Tomita-Takesaki's theory is at the heart of Connes' 1973 classification of
`type III' factors in von Neumann algebras , and on the other hand it has
been put in relation to the KMS condition describing quantum equilibrium
states by Takesaki in his 1970 lecture notes \citep{takesaki1970lnm}, and by
himself and Winnink in 1973, while studying the algebraic properties of KMS
states \citep{takesaki1973cmp}. While the second aspect is the one that is
more relevant for this review, let us mention nonetheless that all
relativistic local algebras of observables in QFT turn out to be type III
factors, in particular of the so-called `type $\mathrm{III}_1$' in Connes'
terminology. They are characterized by the fact that the spectrum of the
modular operator $\Delta$ is the whole real line
\citep[see][]{yngvason2005rmp}. Before Connes' work, inspired as we said by
Tomita-Takesaki theory, the type III factors where very poorly understood (in
fact, the type $\mathrm{III}_1$ was not even defined), and thus this is
another key application of such theory to quantum physics, even if it lies
beyond the scope of this review.

To introduce Tomita-Takesaki's modular operators, and their connection to
equilibrium states in quantum statistical mechanics, let us fix some
mathematical terminology. Preliminarily, let us remark that Tomita-Takesaki's
theory can be developed in a more general framework than the one presented
below, however the setting given here is general enough for the physical
applications we have in mind. Let $\mathscr{H}$ be a separable Hilbert space,
and let $\mathcal{M}$ be a collection of bounded operators (quantum
observables). Then $\mathcal{M}$ is an algebra iff it is closed under
addition and multiplication (\emph{i.e.}, given any $A,B\in \mathcal{M}$, then
$A+B\in \mathcal{M}$ and $AB\in \mathcal{M}$). The \emph{commutant}
$\mathcal{M}'$ of an algebra of operators $\mathcal{M}$ is the collection of
all bounded operators on $\mathscr{H}$ that commute with all elements of
$\mathcal{M}$. Clearly, $\mathcal{M}\subseteq \mathcal{M}''$ (an algebra is contained
in its \emph{bicommutant}). An algebra of operators $\mathcal{M}$ is a
\emph{von-Neumann algebra} iff
\begin{gather*}
  \mathcal{M}=\mathcal{M}''\;.
\end{gather*}
Given a von-Neumann algebra $\mathcal{M}$, the \emph{center} is given by
$\mathcal{M}\cap \mathcal{M}'$. Then, $\mathcal{M}$ is a \emph{factor} iff
$\mathcal{M}\cap \mathcal{M}'=\mathbb{C}\mathds{1}$, \emph{i.e.} the only operators in
$\mathcal{M}$ that commute with all the others are multiples of the identity
operator. Now, let $\Omega\in \mathscr{H}$ be a vector (associated to the quantum
state $\lvert \Omega \rangle\langle \Omega \rvert$); then $\Omega$ is \emph{cyclic and separating} iff both
$\mathcal{M}\Omega$ and $\mathcal{M}'\Omega$ are dense subsets of $\mathscr{H}$. The
main idea behind Tomita-Takesaki's theory is that one can define the
closure of the conjugation antilinear operators acting as
\begin{gather*}
  S\mathcal{M}\Omega=\mathcal{M}^{\dagger}\Omega\;,\\
  F\mathcal{M}'\Omega=(\mathcal{M}')^{\dagger}\Omega\;.
\end{gather*}
They are both densely defined, since $\Omega$ is cyclic and separating. Tomita
proved that their closures exist, they are one adjoint to another
($F=S^{\dagger}$), and they can be expressed, as every closed operator, through an antilinear
polar decomposition $S=J\Delta^{1/2}$, that introduces the modular
operators. The term `modular' is used exploiting the analogy between the operator polar decomposition and the polar expression of complex numbers ($\Delta^{1/2}$ plays the role of the modulus, and $J$ that of the phase, that in addition takes antilinearity into account). Because $S$ and $S^{\dagger}$ conjugate the observables in $\mathcal{M}$
and in its commutant, the modular operators inherit very useful properties,
described below. 

\subsubsection{The modular operator and the modular conjugation}
\label{sec:modul-oper-modul}

Given a couple $(\mathcal{M},\Omega)$, where $\mathcal{M}$ is a von-Neumann
algebra of operators on $\mathscr{H}$ and $\Omega\in \mathscr{H}$ is cyclic and
separating, a couple of operators $(\Delta,J)$ on $\mathscr{H}$ can be defined,
called respectively the \emph{modular operator} and the \emph{modular
  conjugation}. The couple $(\Delta,J)$ is identified uniquely by the following
properties:
\begin{itemize}
\item $\Delta$ is self-adjoint, positive, and invertible; in addition, the vector
  $\Omega$ is invariant under the action of both $\Delta$ and $J$:
  \begin{gather*}
    \Delta\Omega=J\Omega=\Omega\;.
  \end{gather*}
  
\item The unitary group $(\Delta^{it})_{t\in \mathbb{R}}$ defines a \emph{group of automorphisms}
  of the algebra $\mathcal{M}$: for any $t\in \mathbb{R}$,
  \begin{gather*}
    \Delta^{it}\mathcal{M} \Delta^{-it}= \mathcal{M}\;.
  \end{gather*}
  Let us denote by
  \begin{gather*}
    \sigma_t(A)= \Delta^{it}A \Delta^{-it}\;,\qquad A\in \mathcal{M}
  \end{gather*}
  such group of automorphisms, also called the \emph{modular group}.
  
\item $\mathcal{M}\Omega\subseteq D(\Delta^{1/2})$, the latter being the domain of
  self-adjointness of $\Delta^{1/2}$.
  
\item The operator $J$ is a conjugation, \emph{i.e.} $J$ is antilinear,
  $J=J^{\dagger}=J^{-1}$, and it commutes with the modular group: $[J,\Delta^{it}]=0$. Hence,
  \begin{gather*}
    J\Delta J =\Delta^{-1}\;.
  \end{gather*}
  
\item The modular conjugation maps the algebra $\mathcal{M}$ into its
  commutant:
  \begin{gather*}
    J\mathcal{M}J=\mathcal{M}'\;.
  \end{gather*}
  
\item The operators $S=J\Delta^{1/2}$ and $S^{\dagger}=\Delta^{1/2}J=J\Delta^{-1/2}$ `conjugate'
  $\mathcal{M}\Omega$: $\forall A\in \mathcal{M}$, $\forall A'\in \mathcal{M}'$,
  \begin{gather*}
    SA\Omega=A^{\dagger}\Omega\;,\\
    S^{\dagger}A'\Omega= (A')^{\dagger}\Omega\;.
  \end{gather*}
  
\item The function
  \begin{gather*}
    \mathbb{R}\ni t\mapsto \Delta^{it} A\Omega\in \mathscr{H} \;,\qquad A\in \mathcal{M} 
  \end{gather*}
  has an analytic continuation in the strip $\{z\in \mathbb{C}\,,\, -\frac{1}{2}<\Im z <
  0\}$. The function
  \begin{gather*}
    \mathbb{R}\ni t\mapsto \Delta^{it} A'\Omega\in \mathscr{H} \;,\qquad A'\in \mathcal{M}'
  \end{gather*}
  has an analytic continuation in the strip $\{z\in \mathbb{C}\,,\, 0<\Im z <
  \frac{1}{2}\}$. As a consequence, $\forall A\in \mathcal{M}$ and $\forall A'\in
  \mathcal{M}'$,
  \begin{gather*}
    \Delta^{i(t-\frac{i}{2})}A\Omega=\Delta^{it}J A^{\dagger}\Omega\;,\; \Delta^{i(t+\frac{i}{2})}A'\Omega=\Delta^{it}J (A')^{\dagger}\Omega\;.
  \end{gather*}
  
\item The function $t\mapsto \langle \Omega\lvert B\sigma_t(A)\Omega \rangle_{\mathscr{H}}$, $A,B\in \mathcal{M}$,
  can be analytically continued in the strip $\{z\in \mathbb{C}\,,\, -1<\Im z <
  0\}$. At the lower boundary, the following condition holds:
  \begin{equation}
    \label{eq:1MF}
    \langle \Omega\lvert B\sigma_{t-i}(A)\Omega \rangle_{\mathscr{H}}=\langle \Omega\lvert \sigma_t(A)B\Omega \rangle_{\mathscr{H}}\;;
  \end{equation}
  or equivalently,
  \begin{gather*}
    \langle \Omega\lvert B\Delta^{i(t-i)}A \Omega \rangle_{\mathscr{H}}=\langle \Omega\lvert A\Delta^{-it} B\Omega \rangle_{\mathscr{H}}\;.
  \end{gather*}
\end{itemize}
The last property, \cref{eq:1MF}, is called \emph{$\sigma$-KMS condition} for the
state $\lvert \Omega \rangle \langle \Omega \rvert$, and it is related to the notion of equilibrium in
quantum statistical mechanics, as we will explain shortly. Let us remark
that, even if it may not be apparently so, this KMS condition is by itself
\emph{sufficient} to identify the modular couple $(\Delta,J)$.

\subsubsection{Equilibrium states, and their relation to modular groups}
\label{sec:equil-stat-their}

What is an equilibrium state in a quantum theory? Given the evolution of a
physical system as a unitary operator $e^{-itH}$ on some Hilbert space, the
equilibrium state $\omega_{\beta}$ at inverse temperature $\beta\in \mathbb{R}$ is typically given by
the Gibbs prescription:
\begin{gather*}
  \omega_{\beta}(\cdot)=\frac{1}{\tr e^{-\beta (H-\mu N)}}\tr(e^{-\beta (H-\mu N)}\,\cdot\,)\; ,  
\end{gather*}
provided that such a state makes sense mathematically ($\tr e^{-\beta (H-\mu N)}$
could not be finite, for example; this typically happens in taking the
thermodynamic limit). A generalization of the notion of Gibbs state, that
survives the thermodynamic limit and is applicable to QFTs, is that of a
\emph{KMS state}\footnote{KMS is the acronym for Kubo, that first introduced condition \eqref{eq:2MF}, and Martin-Schwinger who first used it in connection to thermodynamic Green functions.}. Let $(\tau_t)_{t\in \mathbb{R}}$ be a group of automorphisms of a von
Neumann algebra of observables $\mathcal{M}$. Then a state $\omega_{\beta}$ on
$\mathcal{M}$ is a \emph{$\tau$-KMS state} at inverse temperature $\beta\in \mathbb{R}$ iff for
any $t\in \mathbb{R}$, and for any $A,B\in \mathcal{M}$,
\begin{equation}
  \label{eq:2MF}
  \omega_{\beta}\bigl(\tau_t(A)B\bigr)= \omega_{\beta}\bigl(B\tau_{t+i\beta}(A)\bigr)\;.
\end{equation}
In other words, the state $\omega_{\beta}$ behaves almost like a trace, the correction
to being a trace `measured' by $\tau_{i\beta}$.

If $\tau_t(A)= e^{itH}Ae^{-itH}$, and the Gibbs state can be defined, then the
latter is a $\tau$-KMS state at inverse temperature $\beta$. It can moreover be
shown that in a finite volume, the only KMS states are the Gibbs states
\citep[see, \emph{e.g.},][Ch.\ V]{haag1992tmp}.

Comparing \eqref{eq:1MF} with \eqref{eq:2MF}, it is clear that $\lvert \Omega \rangle \langle \Omega \rvert$
is a $\sigma$-KMS state at inverse temperature $\beta=-1$. It is perhaps convenient to
re-scale temperature to its physical value $\beta$ in the units of our
choosing. First of all, by functional calculus it is possible to rewrite
\begin{gather*}
  \Delta= e^{-K}\;,
\end{gather*}
where $K$ is now called the \emph{modular Hamiltonian}. In \cref{eq:1MF}, let
us make the substitution $t=-\beta^{-1} s$:
\begin{gather*}
  \langle \Omega\lvert B\sigma_{-\beta^{-1} s-i}(A)\Omega \rangle_{\mathscr{H}}=\langle \Omega\lvert \sigma_{-\beta^{-1} s}(A)B\Omega \rangle_{\mathscr{H}}\;.
\end{gather*}
Now, let us define $H=\beta^{-1} K$, obtaining
\begin{gather*}
  \langle \Omega\lvert B\tau_{s+i\beta}(A)\Omega \rangle_{\mathscr{H}}=\langle \Omega\lvert \tau_{s}(A)B\Omega \rangle_{\mathscr{H}}\;,
\end{gather*}
\emph{i.e.}, $\lvert \Omega \rangle \langle \Omega \rvert$ is a $\tau$-KMS state at inverse temperature $\beta$,
with the evolution $\tau$ generated by $H=\beta^{-1} K$.

We can therefore draw the following key conclusion:
\begin{quote}
  {\itshape Any equilibrium state at inverse temperature $\beta$ can be seen as a
    faithful state\footnote{Faithful states are the abstraction of cyclic and
      separating vectors: given a von Neumann algebra, an abstractly defined
      state is \emph{faithful} iff there exists a concrete representation of the
      algebra in which the state is the projection on a cyclic and separating
      vector.} over the algebra $\mathcal{M}$ of observables, whose modular
    group is given by time translations, with a group parameter $s$ related to
    physical time by $t=-\beta s$.}
\end{quote}

We also have a nice converse piece of information, due to Takesaki
\citep{takesaki1970lnm}. Let $\mathcal{M}$ be a von Neumann algebra, and $\omega$
a faithful state (represented on some $\mathscr{H}$ by $\lvert \Omega \rangle \langle \Omega \rvert$, with
$\Omega$ cyclic and separating), whose associated modular group is given by
$(\sigma_t)_{t\in \mathbb{R}}$. Are there any other equilibrium states, apart from $\omega$, with
respect to the modular group flow? The answer is as follows. Let $\varrho$ be a
density matrix. Then the following two statements are equivalent:
\begin{itemize}
\item $\varrho$ is a $\sigma$-KMS state.
  
\item There exists a (unique) positive operator $T$ on $\mathscr{H}$, whose
  spectral decomposition belongs to the center of $\mathcal{M}$, such that
  for all $A\in \mathcal{M}$,
  \begin{gather*}
    \tr_{\mathscr{H}} \varrho A = \langle T^{1/2}\Omega  , A T^{1/2}\Omega\rangle_{\mathscr{H}}\;.
  \end{gather*}
\end{itemize}

Clearly, if $\mathcal{M}$ is a factor (as many physically interesting
algebras of observables are) then $T=1$, \emph{i.e.} $\omega$ is the \emph{unique}
equilibrium state for the modular group flow.

\subsection{The theorem of Bisognano and Wichmann}
\label{sec:theor-bisogn-wichm}

Bisognano and Wichmann's results, published in a couple of papers between
1975 and 1976 \citep{bisognano1975jmp,bisognano1976jmp}, concern local
algebras of observables in relativistic quantum theories, and in particular
the concept of \emph{Haag duality} for wedge regions, to be explained
below. Their results have then be extended to CFTs by Hislop and Longo
\citep{hislop1982cmp}: in conformal theories, duality can be proved for more
general regions than the wedges considered by Bisognano and Wichmann, in
particular for diamonds (see the illustrative examples in Sec.~\ref{sec:CFTs}).

At a first glance, the result of Bisognano and Wichmann seems unrelated to
Tomita-Takesaki's theory. This is however true only in appearance, in fact
Bisognano and Wichmann themselves remarked the connection of their result to
the modular theory. We focus on such connection, and on the additional
connections with entanglement Hamiltonians, in the last part of this section.
For the moment, let us focus on Bisognano-Wichmann's theorem itself.

\subsubsection{Haag duality in relativistic quantum theories}
\label{sec:haag-duality}

If one wants to build up a (field) theory of relativistic observables in
Minkowski spacetime (or on any other Lorentzian manifold), causality shall be
taken into account. However, there are serious mathematical obstructions to
the definition of pointwise local relativistic quantum observables: as it is
well-known, quantum fields should be smeared by smooth functions supported on
some open region in the Minkowski spacetime, in order for them to make sense
as operators. A field that is smeared by a function supported in a region
$\mathcal{O}$, is said to be localized in $\mathcal{O}$. Causality for
localized fields is assumed in the following form: fields that are localized
on spacelike separated regions of spacetime \emph{shall commute between each
  other}. Algebras of observables in QFT are built by the action of fields
(and possibly their momenta). Given a spacetime region $\mathcal{O}$, the
local algebra of observables $\mathcal{M}(\mathcal{O})$ shall be,
intuitively, the one constructed by the action of fields localized in the
region $\mathcal{O}$. In general, Haag's idea is to start with a collection
of abstractly defined local algebras
$\bigl(\mathcal{M}(\mathcal{O})\bigr)_{\mathcal{O}\subseteq \Sigma}$ of observables, each
defined on an open region $\mathcal{O}$ of a given spacetime $\Sigma$. Such
collection of local algebras shall satisfy certain assumptions that reflect
the physical axioms of a relativistic theory, and in particular causality, in
the form of commutation of spacelike separated observables.

A desirable feature, that is however somewhat difficult to prove, is the
following, called \emph{Haag duality}. Let us denote by $\mathcal{O}'$ the
\emph{causal complement} of the region $\mathcal{O}$, \emph{i.e.} the set of
spacetime points that are at a spacelike distance from all the points of
$\mathcal{O}$. Then duality is true for the region $\mathcal{O}$ iff
\begin{gather*}
  \mathcal{M}(\mathcal{O})' = \mathcal{M}(\mathcal{O}')\;.
\end{gather*}
In other words,
\begin{quote}
  \emph{The commutant of a local algebra of observables in $\mathcal{O}$ consists
  precisely of all observables in the causal complement $\mathcal{O}'$.}
\end{quote}

We will not discuss all the results about duality available in the
literature, let us only mention, in addition to the Bisognano-Wichmann's
papers, the pioneering works by Araki \citep{araki1963jmp2, araki1964jmp},
Dell'Antonio \citep{dellantonio1968cmp}, and Eckmann-Osterwalder
\citep{eckmann1973jfa} for special Minkowski regions and free field's scalar
observables.

\subsubsection{The result of Bisognano and Wichmann: Haag duality on wedges}
\label{sec:result-bisogn-wichm}

Let $\Sigma$ be the Minkowski spacetime: $\Sigma=(\mathbb{R}^4,\eta_{\mu\nu})$, where
\begin{gather*}
  \eta=\begin{pNiceMatrix}[columns-width = auto]
      +&0&0&0\\
      0&-&0&0\\
      0&0&-&0\\
      0&0&0&-
  \end{pNiceMatrix}
\end{gather*}
is the flat pseudo-Riemannian metric. The right wedge $W_{\mathrm{R}}\subset \Sigma$ is
the open set enclosed by two light rays starting from the origin and
propagating in the $x^{1}$ direction:
\begin{gather*}
  W_{\mathrm{R}}=\Bigl\{x=(x^0,x^1,x^2,x^3)\in \Sigma\,,\, x^1>\lvert x^0  \rvert_{}^{}\Bigr\}\;.
\end{gather*}
Its causal complement is the left wedge $W_{\mathrm{L}}$:
\begin{gather*}
  W_{\mathrm{R}}'= W_{\mathrm{L}}=\Bigl\{x=(x^0,x^1,x^2,x^3)\in \Sigma\,,\, x^1<-\lvert x^0  \rvert_{}^{}\Bigr\}\;. 
\end{gather*}

Let us now denote by $\boldsymbol{\Lambda}$ a generic spacetime transformation
belonging to the Poincaré group. Of special usefulness will be the following
Lorentz transformations: the \emph{boosts} $\Lambda(s)$ in the $1$-direction, with
parameter $s\in \mathbb{R}$,
\begin{gather*}
  \Lambda(s)=\begin{pNiceMatrix}[columns-width = auto]
      \cosh s&\sinh s&0&0\\
      \sinh s&\cosh s&0&0\\
      0&0&1&0\\
      0&0&0&1
  \end{pNiceMatrix}\; ;
\end{gather*}
and the \emph{spatial rotation} $R_1(\pi)$, meaning a rotation of spatial
coordinates by an angle of $\pi$ around the $1$-axis.

Let us now consider a possibly interacting quantum field theory on Minkowski
spacetime. In particular, let us suppose that we can properly define the
vacuum representation of the field theory as an algebra of operators acting
on a Hilbert space $\mathscr{H}$ on which the vacuum $\Omega\in \mathscr{H}$ is
represented as a vector, and that there is a unitary representation
$U(\boldsymbol{\Lambda})$ of every Poincaré transformation, suitably acting on
smeared quantum fields, that are defined as self-adjoint operators $\varphi(f)$,
where $f$ is any smooth test function on $\Sigma$. The rigorous construction of
the `correct' local algebras of observables $\mathcal{M}(W_{\mathrm{R}})$
and $\mathcal{M}(W_{\mathrm{L}})$, localized on the right and left wedge
respectively, given by Bisognano and Wichmann is rather technical, and we
shall omit it here. It suffices to keep in mind that these algebras (of
bounded operators) are suitably related to the (unbounded) operators obtained
by forming polynomials of the fields, each one localized in the right or left
wedge, respectively.

The (main) theorem of Bisognano and Wichmann then reads as follows:
\begin{quote}
  \emph{Haag duality holds for the von Neumann algebras
    $\mathcal{M}(W_{\mathrm{R}})$ and $\mathcal{M}(W_{\mathrm{L}})$. More
    precisely (by taking the commutant of the Haag duality for the right
    wedge below, one gets Haag duality for the left wedge):}
  \begin{gather*}
    \mathcal{M}(W_{\mathrm{R}})'=\mathcal{M}(W_{\mathrm{L}})=\mathcal{M}(W_{\mathrm{R}}')\; .
  \end{gather*}
\end{quote}
This result positively resonated in the community of mathematical physics:
the aforementioned duality theorems of Araki and Eckmann-Osterwalder, despite
being far from trivial, were true only for free scalar fields; the results of
Bisognano and Wichmann are true for every field theory that can be reasonably
defined, being it free or interacting. The only drawback, is that one shall
restrict to wedges. As already mentioned, their result (and idea of the
proof) was later extended to conformal fields by Hislop and Longo, where,
thanks to the additional conformal symmetry, more general spacetime regions
such as diamonds could be considered. Bisognano-Wichmann's theorem is still a
topic of research in algebraic QFT nowadays \citep[see, \emph{e.g.},][and
references thereof contained]{morinelli2020cmp, gui2021ahp, neeb2021arx}.

\subsubsection{The link to modular operators}
\label{sec:link-modul-oper}

How is the Haag duality of Bisognano and Wichmann related to
Tomita-Takesaki's modular theory? The link, although not used directly, was
already remarked by Bisognano and Wichmann in their original paper
\citep{bisognano1975jmp}. In fact, \emph{they are able to provide one of the
  very few concrete and explicit realizations of the modular couple
  $(\Delta,J)$}. The special properties of the modular couple, explicit in their
case, play a crucial role in their proof of Haag duality.

The von Neumann algebras they consider are, clearly, the wedge local algebras
$\mathcal{M}(W_{\mathrm{R}})$ and $\mathcal{M}(W_{\mathrm{L}})$. A very
general result of relativistic QFTs, that goes by the name of
\emph{Reeh-Schlieder's theorem} \citep{reeh1961nc,schlieder1965cmp}, ensures
that the vacuum $\Omega$ is cyclic and separating for
$\mathcal{M}(W_{\mathrm{R}})$ (and $\mathcal{M}(W_{\mathrm{L}})$). Hence,
there exists a modular couple $(\Delta_{\mathrm{R}},J_{\mathrm{R}})$ associated to
the vacuum on the right wedge local algebra (and analogously for the left
wedge). To write the couple explicitly, let us consider the unitary
realization of the group of boosts $\Bigl(U\bigl(\Lambda(s)\bigr)\Bigr)_{s\in \mathbb{R}}$. As
any strongly continuous unitary group, it can be written by Stone's theorem
as
\begin{gather*}
  U\bigl(\Lambda(s)\bigr)= e^{isK}\;,
\end{gather*}
with $K$ some self-adjoint operator on $\mathscr{H}$, the quantum generator
of the Lorentz boost. In addition, let us denote by $\Theta$ the CPT-operator on
$\mathscr{H}$, where CPT stands for charge-parity-time reversal (such
operator can always be defined, both abstractly and when possible concretely,
on a relativistic QFT). Then,
\begin{gather*}
  \Delta_{\mathrm{R}}= e^{-2\pi K}\; ,\\
  J_{\mathrm{R}}= \Theta \,U\bigl(R_1(\pi)\bigr)\; .
\end{gather*}
Hence, it also follows that the modular group of the right wedge is given by
the adjoint action of the group of boosts $\Bigl(U\bigl(\Lambda(2\pi
s)\bigr)\Bigr)_{s\in \mathbb{R}}$.

This remarkable aspect of Bisognano-Wichmann's construction is key for
applications: it paved the way to the use of Tomita-Takesaki's theory as the
foundational motivation of important concepts of both theoretical and
experimental physics, in particular that of \emph{Entanglement Hamiltonian}.

\subsection[Modular Hamiltonian and Unruh's effect]{The modular Hamiltonian as an entanglement Hamiltonian: relation
  to Unruh's effect}
\label{sec:modul-hamilt-as}

To explain the connection (even more so, the identification) between modular
and entanglement Hamiltonians, it is useful to reformulate the
Bisognano-Wichmann setting in a slightly different way.

We are given a quantum theory, that is \emph{bipartite}: one part consists of
the right wedge, and the observables $\mathcal{M}(W_{\mathrm{R}})$ localized
on it, the rest of all other observables. The vacuum state $\omega_{\Omega}(= \lvert \Omega \rangle\langle \Omega
\rvert)$ is a ``special state'', being the ground state of the whole system. It is
also a state when restricted to the subsystem of local observables
$\mathcal{M}(W_{\mathrm{R}})$, again playing a special role: it is in fact
faithful, and thus the corresponding modular couple
$(\Delta_{\mathrm{R}},J_{\mathrm{R}})$ can be defined.

However, in making the restriction to the wedge and its local observables,
the nature of the vacuum state changes. In fact, as we explained in
\cref{sec:equil-stat-their}, $\omega_{\Omega}$ is now a thermal equilibrium state, at
temperature $\beta=-1$, for the modular group of the right wedge, that
corresponds to the adjoint action of the group of boosts $\Bigl(U\bigl(\Lambda(2\pi
s)\bigr)\Bigr)_{s\in \mathbb{R}}$, as discussed above. Equivalently, defining the
\emph{entanglement Hamiltonian} as $H=\beta^{-1} K$, and time as $t=- (2\pi \beta)s$,
one obtains that \emph{$\omega_{\Omega}$ is an equilibrium state at temperature $\beta$ for
  the entanglement Hamiltonian time flow}. We can rephrase the last statement
in a more physical fashion:
\begin{quote}
  \emph{In a relativistic field theory, tracing out the degrees of freedom
    other than the ones localized on the right wedge, \emph{thermalizes the
      vacuum}. The resulting state is an equilibrium (Gibbs) state, at
    inverse temperature $\beta<\infty$, with respect to the flow generated by the
    entanglement Hamiltonian $H$, given by a suitable rescaling of the
    generator of Lorentz boosts $K$.}
\end{quote}

The general physical definition of an entanglement Hamiltonian is indeed the
Hamiltonian (in a subsystem) with respect to which the partial trace (with
respect to the other subsystem) of a ground or equilibrium state on the total
bipartite system thermalizes. So, the entanglement Hamiltonian and the
modular Hamiltonian of a von Neumann sub-algebra represent \emph{the same
  physical concept} (provided that the considered state is faithful on the
sub-algebra, but that is mostly a mathematical nuisance).

In a very powerful and almost unique way, Bisognano-Wichmann's theorem
provides an explicit form for the entanglement Hamiltonian on a Minkowski
wedge, for the vacuum of any relativistic field theory.

Let us conclude by remarking that the thermalization of the vacuum in
Bisognano-Wichmann is intimately related (essentially, a mathematical proof)
to Unruh's effect. In fact, the trajectory of the spacetime point
$(0,a^{-1},0,0)$ under the boosts $\Lambda(s)$ is that of an uniformly accelerated
motion on the right wedge, with acceleration $a$. For an observer on this
trajectory that uses his proper time $\tau=\frac{s}{a}$ as the time coordinate,
the generator of time translations, in the coordinate system at rest with
her/him, is none other than $H=a K$. Therefore, by Bisognano-Wichmann's
theorem, the vacuum state $\omega_{\Omega}$ is thermalized for her/him, with
temperature
\begin{gather*}
  T= \frac{a}{2\pi}\; ,
\end{gather*}
that, up to restoring the physical constants always omitted by
mathematicians, is exactly the Hawking-Unruh temperature. In his book
\citep{haag1992tmp}, Haag goes even further by arguing that the right wedge,
populated by observers that are constrained to never leave it, provides the
simplest example of an event's horizon, and this is why Unruh's temperature
coincides with Hawking's temperature. We merely report this interpretation as
a nice little extra feature, to conclude this section, and hopefully to
further invite the readers to the exploration of Tomita-Takesaki's and
Bisognano-Wichmann's beautiful mathematical works.

\section[Integrable models]{Entanglement Hamiltonians of integrable models}\label{sec:int}

In the previous section we took a detour in algebraic field theory in order to provide a strict mathematical background to the BW result and its CFT generalizations in sec. 2. These are the most important examples where the EH of a system composed by continuous degrees of freedom is exactly tractable. From here on we shall rather turn our attention towards particular lattice models, whose low-energy behaviour is known to be described by a relativistic field theory. Then the most important question is how the results derived for the EH in the QFT context generalize to these lattice systems. Indeed, the presence of the lattice breaks the Lorentz invariance of
the theory, which is central to the BW theorem. Nevertheless,
the continuum results turn out to provide, after a proper discretization, a very accurate
description of the lattice EH. In this regard integrable systems play an important role,
since they allow for explicit analytical results. These are available for two different
subsystem geometries, namely a half-infinite chain corresponding to the BW setting,
as well as a finite interval.

\subsection{Half-chain}

The reduced density matrix (RDM) of a half-chain 
$\rho \propto \exp(-\mathcal{H}_{\textrm{\tiny half}})$ can be related to the corner transfer
matrix (CTM) of a corresponding two-dimensional statistical physics model \cite{PKL99}.
The CTM was introduced and studied for integrable models by Baxter \cite{Baxter76,Baxter77,baxterbook},
showing that they allow for an explicit analytical treatment in the thermodynamic limit. 
We first consider the ground state of the transverse Ising (TI) chain
\eq{
\hat H = - \sum_{n=-\infty}^{\infty} \left( \lambda \, \sigma^x_{n} \sigma^x_{n+1} + h \, \sigma^z_{n} \right)
}
where the half-chain RDM is related to the CTM of the two-dimensional Ising model \cite{PKL99}.
The EH has different forms in the ordered ($h<\lambda$) and the disordered phase ($h<\lambda$)
of the chain and reads \cite{Davies88,TruongPeschel89}
\eq{
\mathcal{H}_{\textrm{\tiny half}} =
\begin{cases}
-2I(k') \sum_{n=1}^{\infty} \Big [ \, n \,\sigma^x_{n} \sigma^x_{n+1}
+ k \, \Big(n-\frac{1}{2}\Big) \, \sigma^z_{n}\Big ], & \text{ordered} \\
-2I(k') \sum_{n=1}^{\infty} \Big [ \, k \, n \,\sigma^x_{n} \sigma^x_{n+1}
+ \Big(n-\frac{1}{2}\Big) \, \sigma^z_{n} \Big ], & \text{disordered}
\end{cases}
\label{ehtihc}
}
where $I(k')$ is the complete elliptic integral of the first kind, and the
elliptic parameters are defined as $k=\min(\lambda/h,h/\lambda)$ and $k'=\sqrt{1-k^2}$.
Hence $\mathcal{H}_{\textrm{\tiny half}}$
has exactly the BW form, the energy density being multiplied with a linear term,
albeit with a prefactor that depends explicitly on the ratio $h/\lambda$.
The EH can also be diagonalized exactly and has an equidistant single-particle entanglement spectrum \cite{PKL99}
\eq{
\varepsilon_l =
\begin{cases}
2l \, \varepsilon & \text{ordered}
\\
(2l+1) \, \varepsilon & \text{disordered} 
\end{cases}, \qquad
\varepsilon = \pi \frac{I(k')}{I(k)}
\label{epsl}}
with $l=0,1,\dots$ and spacing $\varepsilon$. Similar results are found for
the anisotropic XY chain, where the RDM is related to the CTM of a triangular Ising model \cite{Peschel04b}.

The CTM approach can also be applied to the interacting XXZ chain
\eq{
\hat H = \sum_{n=-\infty}^{\infty}
\left( \sigma^x_{n} \sigma^x_{n+1} + \sigma^y_{n} \sigma^y_{n+1} +
\Delta \sigma^z_{n} \sigma^z_{n+1}\right),
}
 in the gapped phase ($\Delta>1$), where the half-chain RDM is related to the CTM of the
 six-vertex model \cite{PKL99}, which was studied previously in \cite{baxterbook,Davies89,Frahm91}.
This has again a lattice BW form with
\eq{
\mathcal{H}_{\textrm{\tiny half}} = c(\Delta) \sum_{n=1}^{\infty}
n \left( \sigma^x_{n} \sigma^x_{n+1} + \sigma^y_{n} \sigma^y_{n+1} +
\Delta \sigma^z_{n} \sigma^z_{n+1}\right),
}
where the constant $c(\Delta)$ depends on the anisotropy. The EH can again be diagonalized
exactly, and the spectrum is similar to the TI case \eqref{epsl} in the ordered phase, with
the spacing given by $\varepsilon = \mathrm{arcosh} \, \Delta$. Note that the above results even generalize to the XYZ chain (related to the eight-vertex model) both at the level of the CTM \cite{Thacker86} as well as the corresponding entanglement spectrum \cite{Ercolessi10}.

The CTM method is, however, not restricted to spin chains but allows for
the study of continuous variable systems as well, such as the harmonic oscillator chain
\eq{
\hat H =
\frac{1}{2}\sum_{n=-\infty}^{+\infty}
\left[
p_n^2 + \omega^2 \, q_n^2 +K(q_{n+1} - q_n)^2
\right].
\label{Hhc}
}
Choosing $K=k$ and $\omega=1-k$, the RDM is related to the CTM of the 2D Gaussian model
with an elliptic parametrization of the couplings \cite{PeschelTruong91}. The EH of the
half-chain then reads \cite{PeschelChung99}
\eq{
\mathcal{H}_{\textrm{\tiny half}}
\,=\,
2I(k')
\sum_{n=1}^\infty
 \Big[\,
\Big(n-\frac{1}{2}\Big)\, p_n^2
+
\Big(n-\frac{1}{2}\Big) (1-k)^2\, q_n^2
+
k \, n \, \big(q_{n+1}-q_n\big)^2
\,\Big],
\label{ehoschc}}
which has again a BW form, with single-particle spectrum as for the TI chain in \eqref{epsl} in
the disordered phase. Note that, via a canonical transformation of the positions and momenta,
one can also obtain the result for the parametrization $K=1$ and arbitrary $\omega$.
In this case the elliptic parameter $k$ is the solution of $\omega^2=(1-k)^2/k$ and the prefactor of the EH
reads $2I(k')\sqrt{k}$ \cite{EDGTP20}.

\subsection{Interval}

The case of an interval $A=\left[1,N\right]$ is more difficult to handle,
and a direct calculation of the EH is only possible for free lattice models.
In particular, analytical results are available for an infinite hopping chain
\eq{
\hat H = 
- \sum_n t \, (c^{\dag}_n c_{n+1} + c^{\dag}_{n+1} c_{n})
+\sum_n d \, c^{\dag}_n c_n \, ,
\label{Hff}}
where $c_n^\dag$ and $c_n$ are fermion creation/annihilation operators.
Setting $t=1/2$ and $d=\cos q_F$, the Hamiltonian \eqref{Hff} is diagonalized by a Fourier transform
and the ground state is a Fermi sea with occupied momenta $q\in \left[-q_F,q_F\right]$.
Due to Wick's theorem, the reduced density matrix is given by
$\rho_A = Z^{-1}\exp (-\mathcal{H})$, where the EH is another free-fermion operator
\cite{ChungPeschel01,Peschel03,PE09}
\eq{
\mathcal{H}=  \sum_{i,j=1}^N \, H_{i,j} c^{\dag}_i c_j \, .
\label{EHff}
}
The matrix elements are given by
\eq{
H_{i,j}= \sum_{k=1}^N\,\phi_k(i)\; \varepsilon_k\; \phi_k(j) \, , \qquad
\varepsilon_k = \ln \frac {1-\zeta_k}{\zeta_k} \, ,
\label{Hij}}
where $\zeta_k$ and $\phi_k(i)$ are the eigenvalues and eigenvectors 
of the reduced correlation matrix $C_A$, with matrix elements
$C_{i,j}=\langle c_i^\dag c_j \rangle$ restricted to $i,j\in A$.

\begin{figure}[thb]
\centering
\includegraphics[width=0.5\textwidth]{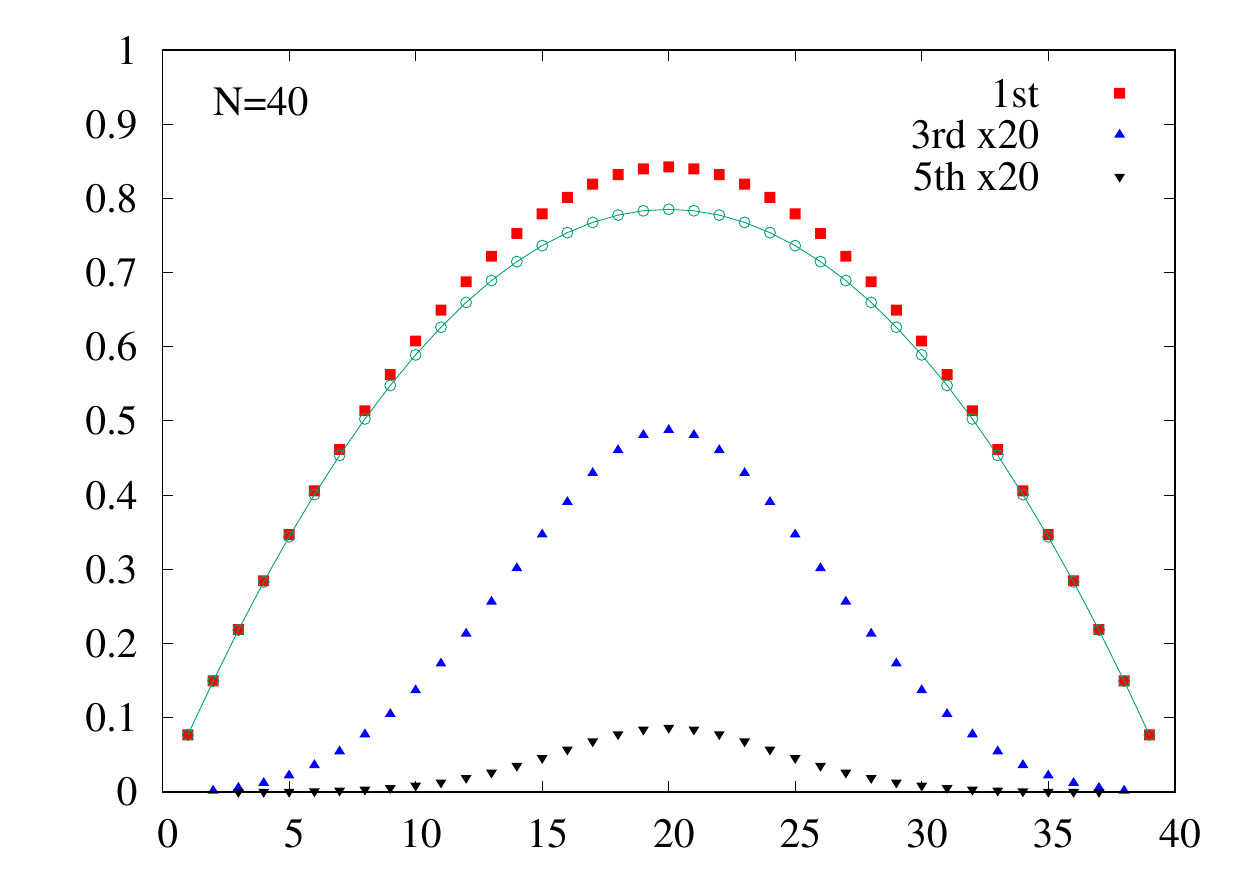}
\caption{Hopping matrix elements in $-H/N$ and nearest-neighbour hopping in $\pi T$ (green) as 
functions of the position for $N=40$ sites at half filling. The 3rd and 5th neighbour hopping
amplitudes are magnified by a factor of $20$ for better visibility. Figure originally published in Ref.~\cite{EP17}.}
\label{fig:EHff}
\end{figure}

The expression \eqref{Hij} allows for a calculation of the EH,
which requires very high precision numerics, as the dominant $\varepsilon_k$
contributions originate from eigenvalues $\zeta_k$ that lie exponentially
close to zero or one. The analytical treatment, however, follows a different
route based on the existence of a commuting tridiagonal operator $T$ \cite{Slepian78,Peschel04},
with matrix elements $T_{i,i}=d_i$ and $T_{i,i+1}=T_{i+1,i}=t_i$ given by
\eq{
t_i = \frac{i}{N} \left(1-\frac{i}{N}\right),\qquad
d_i = -2\cos{q_F}\;\frac{2i-1}{2N}\;\left(1-\frac{2i-1}{2N}\right) .
\label{trid}}
Thus $T$ describes an inhomogeneous hopping chain, with hopping amplitudes
following the exact same parabolic profile as one would obtain from a
proper discretization of the CFT result \eqref{eq:CFT1}. This suggests the relation
$H= -N\pi T$, which was also found for the low lying $\varepsilon_k$ eigenvalues \cite{Peschel04}.
However, comparing to the numerical results, one finds some discrepancies
as shown in Fig.~\ref{fig:EHff} for half filling ($q_F=\pi/2$). Indeed, one observes a slight
deviation from the expected nearest-neighbour hopping profile, as well as nonvanishing
hopping to more distant sites (note that only odd distances appear due to the particle-hole symmetry).

%

The difference can be understood by a proper comparison of the spectra of $h=-H/N$ and $T$,
which was studied by Slepian in \cite{Slepian78}. It turns out that, in the limit $N\to\infty$, one can
relate them via a series expansion
\eq{
h = \sum_{m=0}^{\infty}
\alpha_m \beta_m T^{2m+1} \, ,
\label{ht}}
with appropriate constants $\alpha_m$ and $\beta_m$ \cite{EP17}. 
The leading order has $\alpha_0\beta_0=\pi$ and thus gives the expected CFT behaviour.
However, higher powers of $T$ generate more distant hopping and also modify
the nearest-neighbor one. Interestingly, the matrix elements of $h$ can be found in
a closed form via generalized hypergeometric functions. In particular, 
in the limit $i,N \to \infty$ with $x=i/N$ fixed, one has \cite{EP17}
\eq{
h_{i,i+1} = \pi \, x(1-x) \, 
_{3}F_{2}\left(\frac{1}{4},\frac{1}{2},\frac{3}{4}; 1,2 ; \left[4x(1-x)\right]^2\right)
\label{hnn}}
where the parabolic profile is multiplied by a function which increases smoothly
from 1 at $x=0$ to 1.076 at $x=1/2$ and thus gives a deviation of roughly $8\%$
in the middle. Similar formulae are found for $h_{i,i+r}$ with $r$ odd, showing
a strong suppression of the distant hopping. Nevertheless, the maxima of the
profiles at $x=1/2$ decays only as a power law $r^{-3}$ \cite{EP17}.
Note that the analytical treatment can be extended for general fillings, where
the relation \eqref{ht} between $h$ and $T$ becomes more complicated \cite{EP17},
and also hopping with even $r$ appear.

A similar approach can also be applied for the study of the EH in a \emph{finite} chain.
In fact, a commuting tridiagonal matrix $T$ exists for both periodic \cite{Gruenbaum81}
as well as for open chains \cite{EP18}, describing a hopping profile that is
again the discretized version of the CFT result in the corresponding
geometries \cite{CardyTonni16}. Analogously to the infinite chain case, however,
the nearest-neighbour hopping in the EH is slightly larger around its maximum,
and longer range terms with much smaller amplitudes are also present.
Numerical investigations suggest a relation similar to \eqref{ht}, where the
coefficients depend only on the ratio of subsystem and full system lengths,
but their analytical expression is not known explicitly \cite{EP18}.
Note that, based on the theory of bispectrality, commuting tridiagonal matrices
have also been identified for particular \emph{inhomogeneous}
free-fermion Hamiltonians \cite{CNV19,CNV21}, the corresponding relation \eqref{ht}
was, however, not yet studied.

Finally, we discuss the case of a non-critical chain, obtained by dimerizing the
couplings $t_{2n-1}=(1-\delta)/2$ and $t_{2n}=(1+\delta)/2$ of the hopping chain
in \eqref{Hff}. Using the spin-representation and introducing dual variables
\cite{PerkCapel77,PeschelSchotte84,Turban84,IgloiJuhasz08},
such a Hamiltonian can be mapped into two interlacing TI chains where
$h_1=t_{2n-1}$ and $\lambda_1=t_{2n}$, whereas the couplings
in the second chain are interchanged. For a half-chain, this would yield
an EH of the form \eqref{ehtihc}, where the two terms are now replaced by
the even and odd hopping terms, with the elliptic parameter $k=(1-\delta)/(1+\delta)$
multiplying the odd ones for $\delta>0$.
One can now argue that, if the size of the interval is much larger then the
correlation length ($N \gg \xi$), then the contributions from the two boundaries should
decouple in the RDM. Thus the EH should effectively behave as a half-infinite chain
from each end, with the profile given by
\eq{
h_{2i-1,2i} \simeq 2I(k') \, k \, \Delta((2i-1)/N)\,,
\qquad
h_{2i,2i+1} \simeq 2I(k')\, \Delta(2i/N)
\label{hdim}}
with the \emph{triangular} function
\eq{
\Delta(x)=
\Bigg\{\begin{array}{ll}
x & 0\leq x \leq 1/2 \\
1-x \hspace{.6cm}& 1/2 \leq x  \leq 1 \\
\end{array}.
\label{triangle}}

\begin{figure}[thb]
\center
\includegraphics[width=0.5\textwidth]{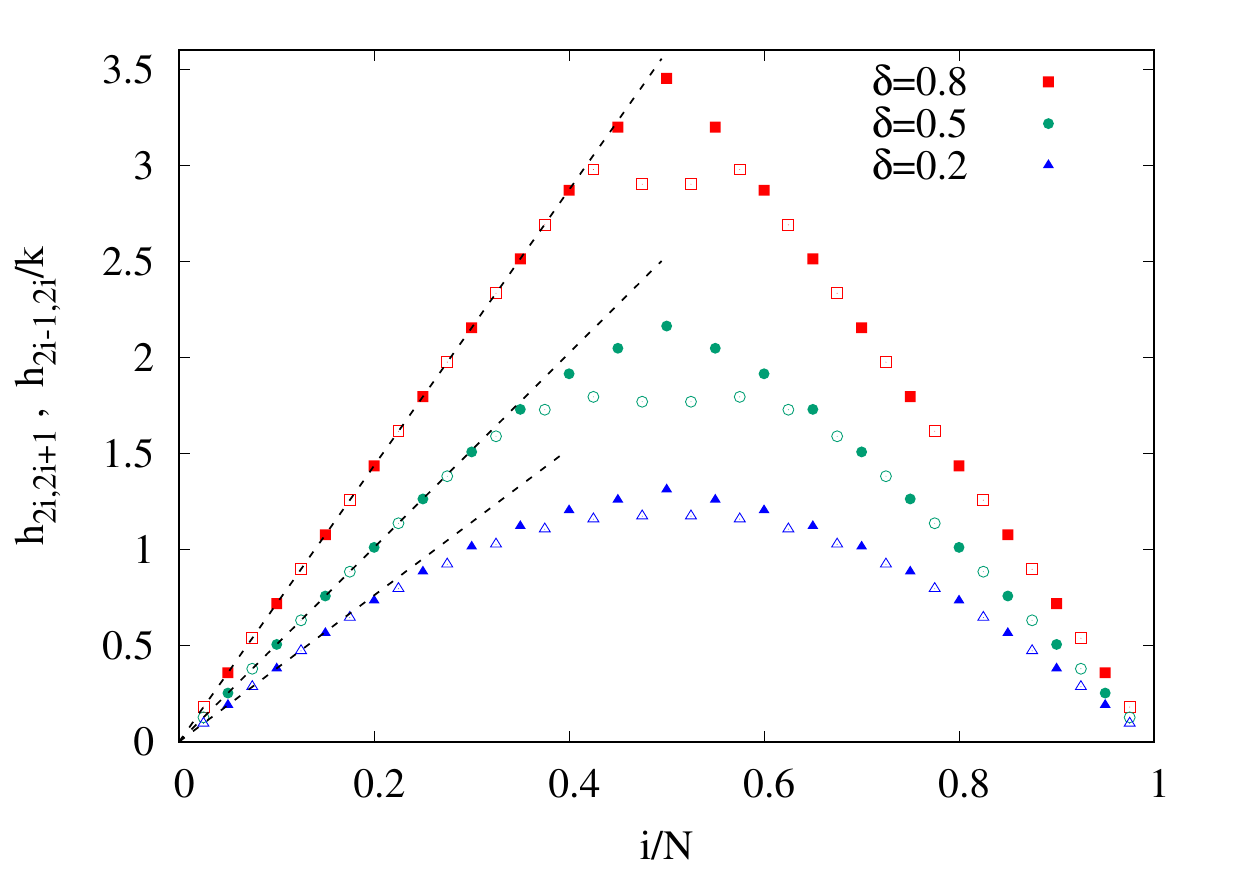}
\caption{Nearest-neighbour hopping in the EH of the dimerized chain
for various $\delta$ and $N=40$. The hopping across odd bonds (empty symbols) are divided by a factor
of $k$. The dashed lines have slopes $2I(k')$ corresponding to the result \eqref{hdim}. Figure originally published in Ref.~\cite{EDGTP20}.}
\label{fig:hnndim}
\end{figure}

The numerical results are shown in Fig.~\ref{fig:hnndim}, and are well described by 
\eqref{hdim} in the vicinity of the interval edges, with some deviations around the center.
The triangular approximation improves with decreasing correlation length $\delta \to 1$, however,
towards the critical limit $\delta \to 0$ one observes a crossover to a roughly parabolic form.
 
%
%

\subsection{Continuum limit}

The exact form of the EH for an interval in a critical free-fermion chain thus shows characteristic deviations
from the continuum CFT results, which need to be properly understood. On one hand, one could
check whether the discretized CFT result (i.e. replacing $\mathcal{H} \to -N \pi T$ in the RDM) would produce an error in the expectation values of local observables that vanishes in the limit $N \to \infty$. Indeed, using this approximation one observes, that the fermionic correlation matrix $C_A$ can be reproduced to very high accuracy even for smaller values of $N$ (although its translational invariance is lost).
Moreover, one can make even stronger statements by considering the trace distance between the two RDMs as a function of $N$, as shown in the next section.

On the other hand, it would be desirable to understand how the proper continuum limit
of the lattice EH arises, when we introduce a lattice spacing $s$ and consider the limit
$N\to \infty$ and $s \to 0$ with $\ell = Ns$ fixed. For the simple homogeneous hopping
chain \eqref{Hff}, the standard procedure is to linearize the dispersion around the Fermi
points $\pm q_F$, and introduce slowly varying fields that describe left- and right-moving
fermions. Then one ends up with a massless Dirac Hamiltonian with a single parameter given by the
Fermi velocity $v_F$. Let us now add long-range hopping over $2p+1$ sites with amplitude $t_{2p+1}$,
which enters the dispersion with a factor $2\cos\left[(2p+1)qs\right]$ and thus modifies $v_F$.
Finally assume, that the argument holds true for hopping amplitudes that vary slowly in space
$t_{2p+1}(x) = N h_{i-p,i+p+1}$ as in the EH, where $x=i s$ is a continuous coordinate.
The space-dependent Fermi velocity  then reads \cite{ETP19}
\eq{
v_F(x) = 2 \ell \sum_{p} (-1)^p (2p+1) h_{i-p,i+p+1} \,,
\label{vfx}}
where the alternating factor is due to $q_F s=\pi/2$ at half filling.
Inserting \eqref{ht}, the resulting double infinite sum \eqref{vfx} can be carried out exactly
and delivers $v_F(x)=2\pi \, x \, (\ell-x)/\ell$, which is identical to the weight function
in the CFT result \eqref{eq:CFT1}.

\begin{figure}[htb]
\center
\includegraphics[width=0.49\textwidth]{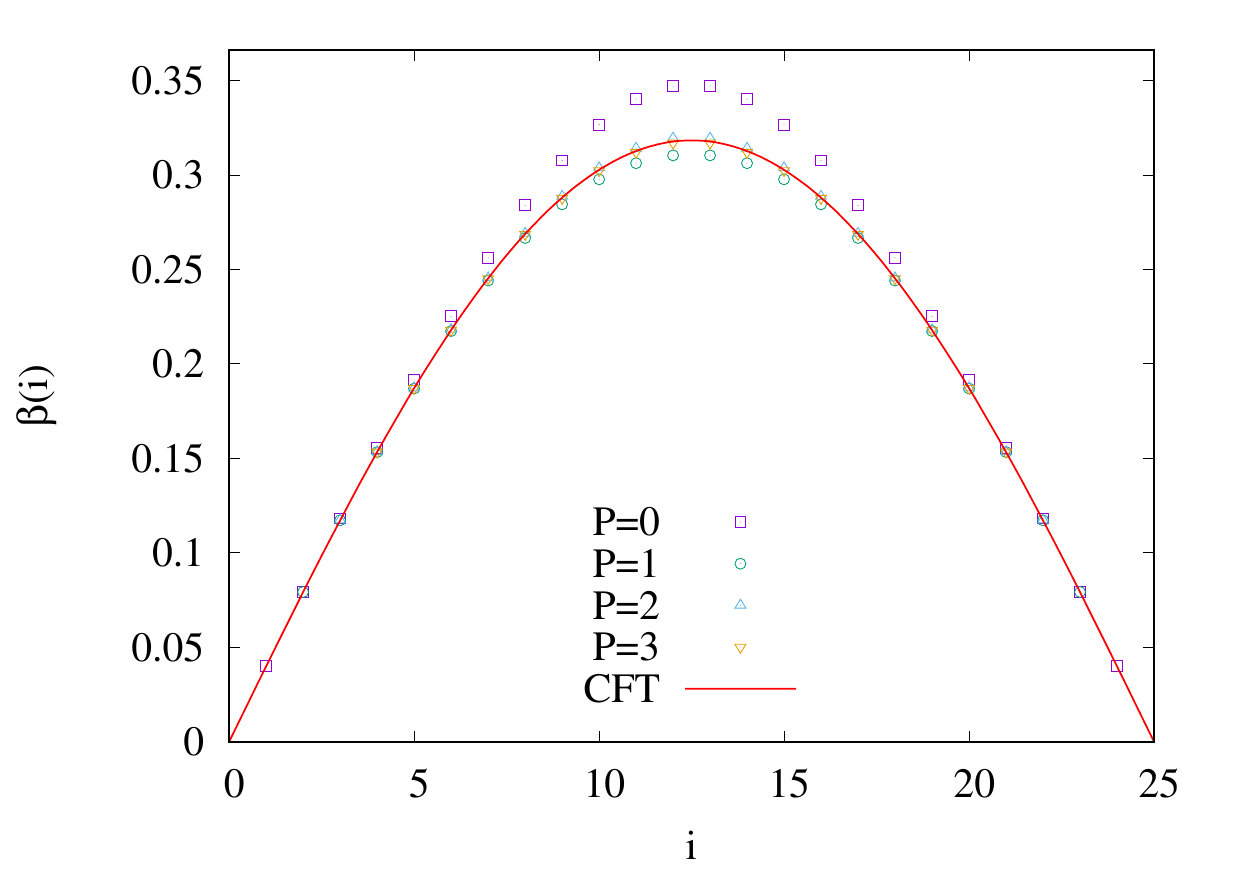}
\includegraphics[width=0.49\textwidth]{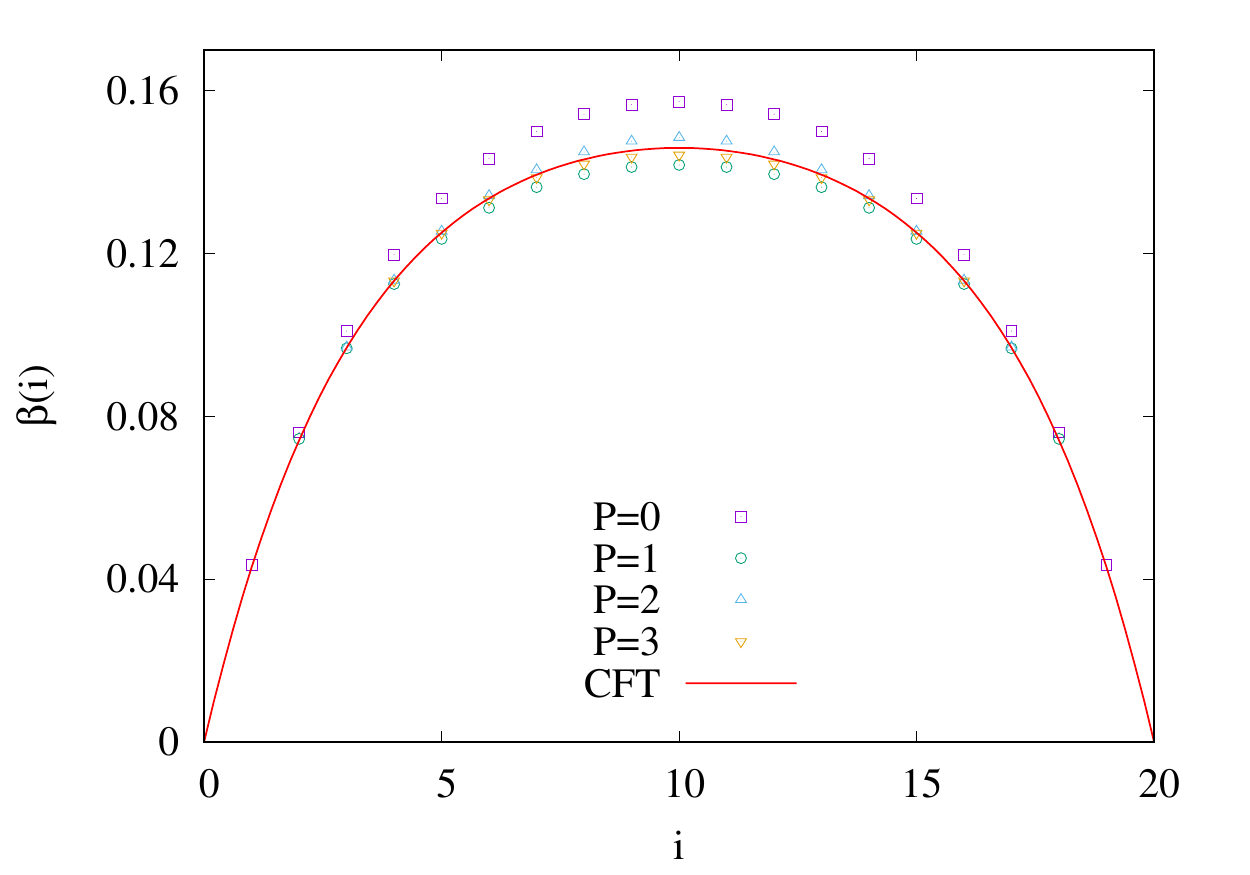}
\caption{Left: $\beta(i)=v_F(x)/2\pi\ell$ for an interval of $N=25$ sites in a finite ring with subsystem
ratio $r=1/2$. The symbols correspond to different cutoffs $P$ of the sum in \eqref{vfx}, while
the solid line shows the CFT result \eqref{eq:CFT2}.
Right: same for an interval of $N=20$ sites in an infinite chain at inverse temperature
$\beta=20$, compared to the CFT result \eqref{eq:CFT4}. Note the different vertical scales. Panels originally published in Ref.~\cite{ETP19}.}
\label{fig:vfx}
\end{figure}

Hence, the function multiplying the energy density in the continuum treatment
has now the interpretation of a local Fermi velocity (instead of a local inverse temperature) that follows from the lattice EH via \eqref{vfx}.
Remarkably, this relation can be generalized to arbitrary fillings 
and one obtains the exact same analytical result for $v_F(x)$ \cite{ETP19}.
The formula \eqref{vfx} can also be used in numerical calculations for chains of finite length
or at finite temperatures \cite{ETP19} (see Fig.~\ref{fig:vfx}), as well as for disjoint segments \cite{ABCH17},
perfectly reproducing the corresponding CFT results \cite{CardyTonni16,CasiniHuerta09}.
Furthermore, the continuum limit of the EH for an interval in the harmonic chain \eqref{Hhc}
can be dealt with along similar lines of thought. This yields again the renormalized couplings
as weighted sums over the diagonals of the lattice EH, perfectly reproducing the CFT
result in the massless limit \cite{DGT20}.
In contrast, in the massive case one experiences convergence issues with the sums \cite{EDGTP20},
which might indicate that the EH becomes non-local even in the continuum limit,
as suggested by a perturbative treatment in the CFT context \cite{ABCH17}.

%
%

\section[Beyond exactly-soluble models]{Entanglement Hamiltonians on the lattice beyond exactly-soluble models}\label{sec:lattice}

The insight from integrable lattice systems is remarkable: the functional form of BW is exactly recovered for massive theories on the half-chain, albeit with a prefactor that depends on the mass term. For gapless theories on the interval, one shall handle with more care: however, there is typically a large overlap between the exact and BW guessed Hamiltonian. 

These works have stimulated the study of generic lattice Bisognano-Wichmann (LBW) entanglement Hamiltonians, to understand whether the latter have predictive power. In this section, we will review the current status of this search. Instead of following a historical perspective (which shall start with Li and Haldane work on topological phases), we prefer to start with critical one-dimensional theories, that bridge more naturally to the previous section, as well as with the CFT predictions in Sec.~\ref{sec:CFTs}. We will then cover massive phases in 1D (including symmetry-protected topological phases), and dimensions larger than one. Before continuing, it is worth mentioning that, from the  perspective of state characterization, entanglement Hamiltonians have also been discussed in the context of representation of tensor network states~\cite{Schollwock2011}. This research line, which has been proved very insightful on its own, is quite distinct from the state characterization of generic Hamiltonian eigenstates that we discuss here, so we refer the reader to specific reviews in field for a detailed discussion - see, in particular, Ref.~\cite{Cirac2021}. 

\subsection{Bisognano-Wichmann theorem on the lattice}

As most field theory results, the Bisognano-Wichmann theorem can be adapted to the lattice via space discretization. This procedure is particularly delicate however, as one is dealing with what is a very complicated operator (the logarithm of the reduced density matrix,  capturing arbitrary correlation functions), whose definition presents challenges close to the boundary. We review here a formulation of such discretization following Ref.~\cite{Giudici2018}, that is convenient for both 1D and 2D systems. 

\begin{figure}[t]
\center
\includegraphics[width=0.49\textwidth]{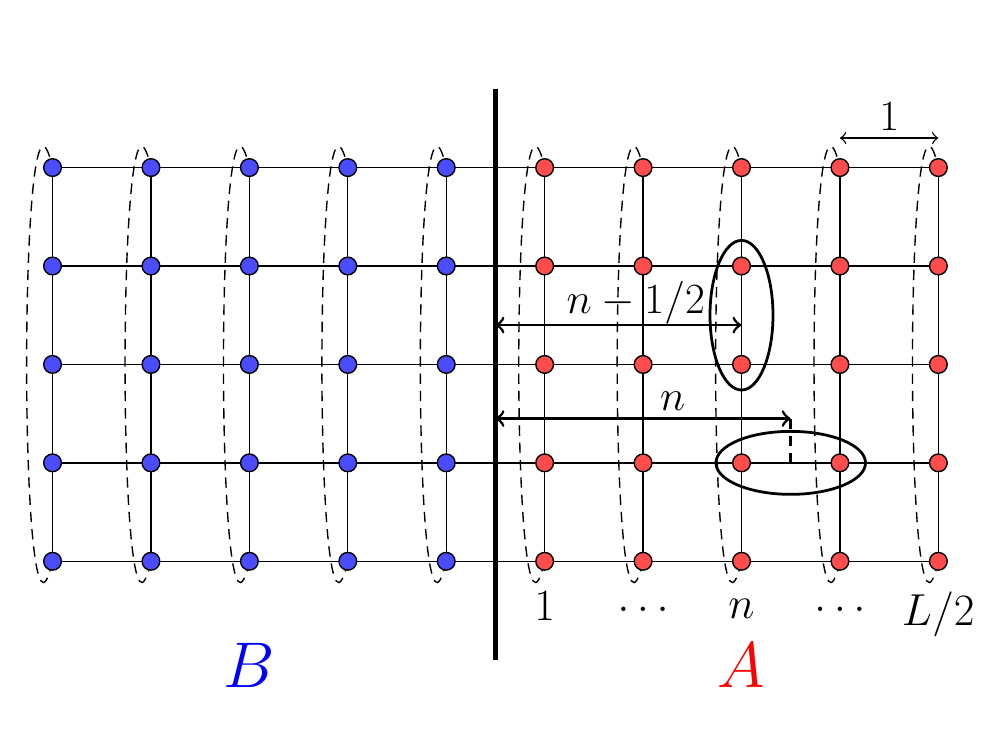}
\caption{Graphical representation of the two-dimensional lattice with partitions $A$ and $B$, and  corresponding notations of the distance from the boundary in units of the lattice spacing: nearest-neighbor terms along the vertical ($y$) direction, and local terms have distance $n-1/2$, while nearest-neighbor terms along the horizontal ($x$) direction have distance $n$. This figure was originally published in Ref.~\cite{Giudici2018}.}
\label{fig:LBW}
\end{figure}

For the sake of simplicity, we consider a square lattice, with a bipartition as shown in Fig.~\ref{fig:LBW}. We are interested in characterizing the ground state wave function of a Hamiltonian that reads as:
\begin{align}
 H =  \Gamma\sum_{x,y,\delta=\pm1} \left[ h_{(x,y),(x+\delta,y)}  +  h_{(x,y),(x,y+\delta)} \right]  + \Theta \sum_{x,y} l_{(x,y)},
 \label{Ham}
\end{align}
where the first two terms represent nearest-neighbor contributions along the $x$ and $y$ direction, respectively, and the last term is onsite. 

Discretizing at first order in the lattice spacing the BW theorem leads to the following lattice BW EH:
\begin{align}
 \mathcal{H}_{A,BW} = \beta_{EH}  \sum_{x,y,\delta=\pm1} \left(\Gamma_x h_{(x,y),(x+\delta,y)}  + \Gamma_y h_{(x,y),(x,y+\delta)} \right)               +  \sum_{x,y} \Theta_{x,y} l_{(x,y)} ,
 \label{BW-EH}
\end{align}
where the inhomogeneous couplings and on-site terms depend on the distance from the  boundary separating subsystem $A$ and $B$ (see Fig.\,\ref{fig:LBW}) 
according to the geometry of the original system, and the inverse entanglement temperature $\beta_{EH}$ includes both constants and the characteristic velocity on the lattice (that corresponds to the speed of sound for critical theories). 

\paragraph{{\it Half-plane lattice BW EH. -}} For the case of a system dynamics that is governed by a generic, relativistic field theory (either massive or not), the BW theorem suggest the following dependence of the EH couplings:
\begin{eqnarray}\label{eq_dist}
\Gamma_x &=& x\Gamma, \nonumber  \\
\Gamma_y &=& \left(x - \frac{1}{2}\right) \Gamma, \nonumber \\
\Theta_{(x,y)}& =& \left(x - \frac{1}{2}\right) \Theta. 
\end{eqnarray}
Note that there is an important point to be emphasized, that is, what is the corresponding distance from the boundaries of the various terms (a feature also encountered in the case of the Ising half-chain \eqref{ehtihc} in the previous section). The rule chosen here is as follows: for local terms, we took as distance from the boundary their shortest euclidean distance. For instance, in the 1D Ising case for $h_{n}$, this corresponds to $n-1/2$. For terms defined on bonds (such as kinetic energy and nearest-neighbor spin couplings), one instead utilizes as a distance from the boundary the shortest euclidean distance from the center of the bond (see Fig.~\ref{fig:LBW}). While strictly speaking applicable to the infinite half plane only, the formulas above are expected to work equally well for other kinds of partitions with a single interface and no angles - such as half-partitions of infinite cylinders. Finally, we note that the same discretization is obtained as a formulation of the lattice Unruh effect~\cite{Celi2017}.

\paragraph{{\it Finite partition in a finite chain. -}} The formulas above, while very generic, are strictly speaking applicable solely to the case of an infinite partition in an infinite system. 
For the case of a (1+1)-d conformal field theory, it is possible to adapt to the lattice the conformal extensions discussed in Sec.~\ref{sec:CFTs}.
In particular, for a periodic chain of length $L$, the half-partition case reads:
\begin{eqnarray}
\Gamma_x &=& \frac{L}{2 \pi} \sin \left( \frac{ 2 \pi x }{ L } \right) \, \Gamma ,\nonumber  \\
\Theta_x &=& \frac{L}{2 \pi} \sin \left(  \frac{ 2 \pi}{L} \left( x - \frac{1}{2} \right) \right) \, \Theta ,
\label{eqdist2}
\end{eqnarray}
while for the open chain
\begin{eqnarray}
\Gamma_x &=& \frac{L}{\pi} \sin \left(  \frac{ \pi x }{ L } \right) \, \Gamma, \nonumber  \\
\Theta_x &=& \frac{L}{\pi} \sin \left(  \frac{ \pi}{L} \left( x - \frac{1}{2} \right) \right) \,  \Theta .
\label{eqdist3}
\end{eqnarray}
To connect the above equations to the CFT expressions in Eqs.~\eqref{eq:CFT2} and \eqref{eq:CFT3}, the corresponding inverse entanglement temperature shall satisfy $\beta_{EH}=\frac{2\pi}{v}$, where $v$ is the speed of sound of the lattice model. Note that $v$ depends on the model parameters and may differ from the maximal group velocity. For instance, in case of free fermions discussed in Sec. \ref{sec:int}, $v$ is simply the Fermi velocity and thus depends on the filling. In principle, the CFT predictions can also be lattice-regularized at finite temperature in the same manner. The main questions is then how well such lattice regularizations approximate the exact EH, which can be investigated numerically.

\subsection[Numerical investigations of lattice Bisognano-Wichmann]{Numerical investigations of lattice Bisognano-Wichmann entanglement Hamiltonians in one-dimensional systems}

Differently form the direct evaluation of the lattice entanglement Hamiltonian available for certain integrable models~\cite{PeschelChung99,Nienhuis_2009}, for non-integrable systems, as well as for interacting systems under arbitrary boundary conditions, investigations of the lattice BW theorem as formulated in the previous paragraphs are typically carried out in a indirect manner. The reason is that a direct extraction of the Hamiltonians terms is a computationally cumbersome operation, that, to date, can only be performed utilizing a specific ansatz for the couplings. At the exact level, this is typically feasible only on relatively small system sizes. Approximate numerical schemes have been devised, utilizing either parent-Hamiltonian-type methods~\cite{PhysRevB.99.235109}, or quantum Monte Carlo sampling~\cite{ParisenToldin2018}. We will comment on those at the end of the next subsection.

The indirect route to study EH of interacting systems relies on validating {\it a posteriori} the lattice BW EH. In fact, what one is really checking is whether, for the ground state of a given model Hamiltonian, the true EH and the lattice BW EH have the same properties up to a given threshold error. Various types of tests have been devised for this purpose. The most immediate choice is to directly compare their spectra and their eigenstates; more 'coarse-grained' tests include entropies (in this context, construed as moments of the distribution of the entanglement spectrum) and expectation values of order parameters and correlation functions. In some cases, it is also possible to compare the operator distance(s) between the EH and the lattice BW EH: those are by far the strictest tests. In the following, we present a few, selected examples that illustrate potentials and challenges of these various approaches. We will denote with $\rho_A$ the system RDM, and with $\rho_{A, BW}$ the one obtained from the lattice BW EH.

\paragraph{Ising models: finite-volumes and long-range interactions. -} We have seen in the previous section that the EH of the Ising model away from its critical point, for the half-chain partition in an infinite chain, is exactly given via the corner transfer matrix. At the critical point and for finite partition sizes, an exact solution is not available. 

\begin{figure}[t]
\center
\includegraphics[width=0.45\textwidth]{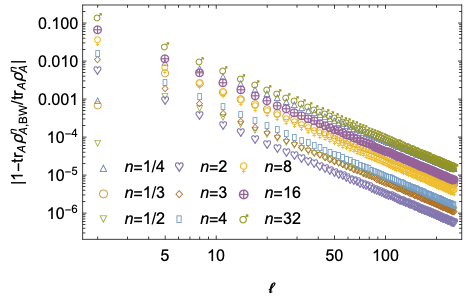}
\includegraphics[width=0.40\textwidth]{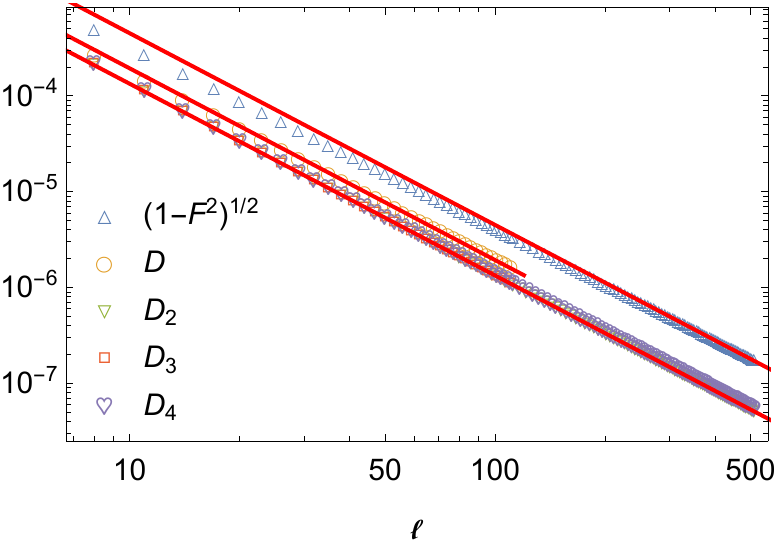}
\caption{Relative difference between the lattice BW EH, $\rho_{A,BW}$, and the microscopic RDM $\rho_A$, for a finite partition $A$ of length $\ell$ in the Ising model at the critical point.   Left panel (interval in a infinite chain): difference between $n$-th moments of the RDM. Right panel (interval in a finite chain): decay of different distances ($F$: Uhlmann fidelity; $D_n$: normalized $n$-distances) between the lattice BW EH and the exact EH. Red lines are proportional to $\ell^{-2}$. The two panels of this figure were published in Ref.~\cite{Zhang2020}.}
\label{fig:IsingPBC}
\end{figure}

In Ref.~\cite{Giudici2018,TMS2019,Zhang2020}, a systematic comparison between the lattice BW EH and the real EH was performed, at the level of both eigenvalues, eigenvectors, and full RDM. In Fig.~\ref{fig:IsingPBC}, some of the properties of the RDM are considered, for partition sizes $\ell$ embedded in an infinite chain (left), or finite periodic chain (right) of length $L=4\ell$. The left panel shows how different moments of the entanglement spectrum distribution (that is, the traces of the RDM to a given power) become closer and closer as the partition size is increased. This signals that the entanglement spectrum of the lattice BW EH is approaching the same distribution of the real ES. 

In the right panel, we show the scaling of several operator distances between the real EH, and the lattice BW one, from Ref.~\cite{Zhang2020}. In particular, $D_n$ are generalized Schatten distances:
\begin{equation}
    D_n = \frac{1}{2^{1/n}}||\rho_{A,BW}-\rho_A ||_n,
\end{equation}
$D=D_1$ is the trace distance, and the triangles are related to the Uhlmann fidelity $F = \text{tr}\sqrt{\sqrt{\rho_{A, BW}}\rho_A\sqrt{\rho_{A, BW}}}$. All distances scale systematically to zero in the thermodynamic limit. This signals the fact that, while at finite size there are discrepancies between the real and lattice BW EH, the corresponding RDMs converge to the same operator in the limit of large enough partitions. Similar results have been obtained for XY spin chains.   

These type of accurate verification can only be performed for quadratic Hamiltonians: it is, for instance, not feasible for the (experimentally relevant) case of long-range Ising models. Still, the latter entanglement spectra have been shown to be accurately reproduced by the lattice BW EH, at least in cases where long-range interactions decay as dipolar or van der Waals cases~\cite{Dalmonte2018}. 

\begin{figure}[t]
\center
\includegraphics[width=0.40\textwidth]{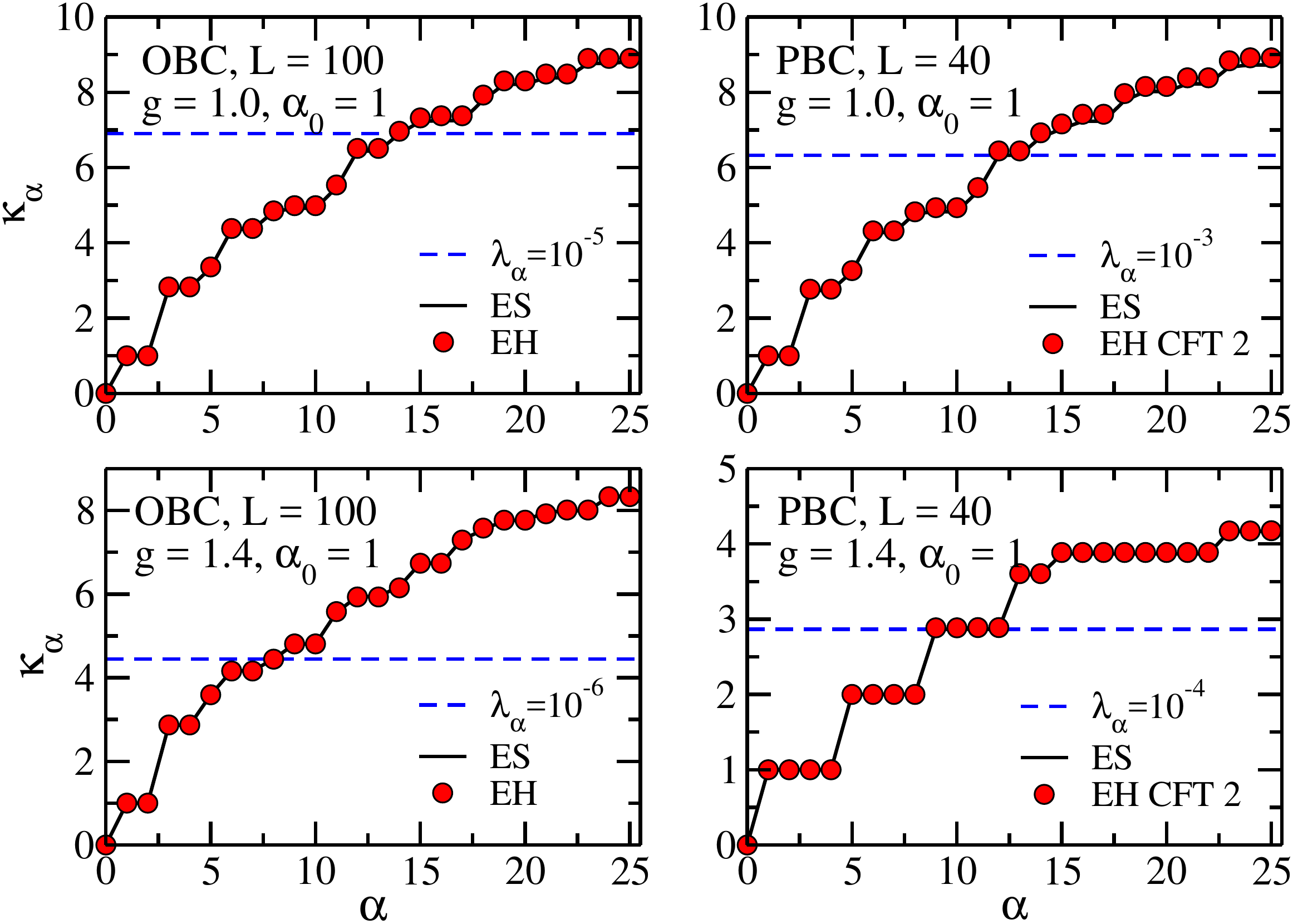}
\includegraphics[width=0.40\textwidth]{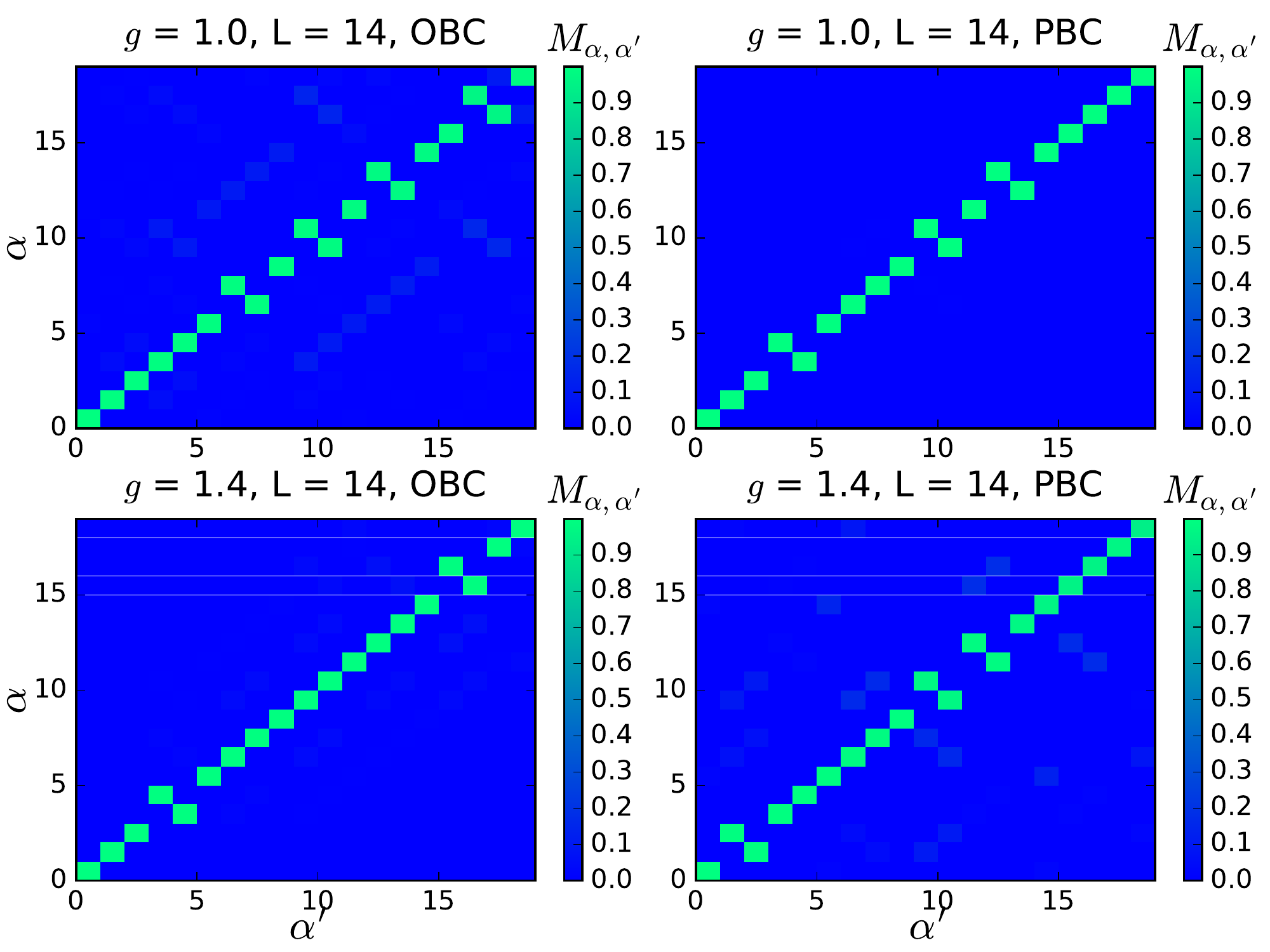}

\caption{Three-state Potts model. Left panels: comparison between the entanglement spectra of the microscopic model (black lines) and the ones obtained from the lattive BW EH (red dots). Right panels: overlaps between entanglement eigenvectors. The panels of this figure are taken from Ref.~\cite{Giudici2018}.}
\label{fig:Potts}
\end{figure}

\paragraph{Potts models. -} For generic quantum systems, different ways of comparing $\rho_A$ and $\rho_{A,BW}$ have been introduced. One practical method for numerical simulations is to compare their eigendecompositions, both in terms of spectrum and vectors. As far as the former is concerned, the following ratios are typically employed~\cite{Dalmonte2018}:
\begin{equation}
    \kappa_\alpha = \frac{\epsilon_\alpha-\epsilon_0}{\epsilon_1-\epsilon_0}
\end{equation}
where $\epsilon_\alpha$ are the entanglement eigenergies in increasing order. Such ratios have the practical advantage of being insensitive to the definition of entanglement temperature, as well as to additive constants in $\mathcal{H}_A$. In Fig.~\ref{fig:Potts}, first two columns, we show such ratios corresponding to $\rho_A$ (black line) and $\rho_{A, BW}$ (red dots) for the case of the one-dimensional three-state Potts model. For both OBC and PBC, as well as at ($g=1.0$) or away from ($g=1.4$) the critical point, the lattive BW EH is able to accurately predict the full low-lying spectrum down to values smaller than $10^{-6}$ even for the modest system sizes considered here (the partition $A$ is of length $L/2$). Note that the several degeneracies present outside of the critical point (bottom row) are captured at the percent level.

Eigenvectors can also be compared one-by-one, at least for small system sizes where a full diagonalization of the lattice BW EH can be performed. In the right panels of Fig.~\ref{fig:Potts}, we show such comparison for a system of size $L=14$, where the quantity plotted is:
\begin{equation}
    M_{\alpha,\alpha'} = |\langle\psi_{A;\alpha}|\psi_{A, BW; \alpha'}\rangle|.
\end{equation}
Deviations from zero outside of the diagonal are smaller than $10^{-3}$, while degeneracies in the spectrum appear as finite terms just next to the diagonals (both for $g=1.0, 1.4$). Overall, the agreement is excellent even at such small sizes.

\paragraph{Other critical chains. -} The analysis above has been performed for a broad variety of critical points and phases, including Heisenberg chains, spin-1 bilinear-biquadratic, XX and XY models, and Hubbard models. Overall, the results support the picture that is also suggested by the exactly soluble models discussed in the previous sections: when not exact, the lattice BW EH often provides a very accurate description of the real EH, both in terms of its generic properties (e.g., locality and types of terms) and of its spectral properties (eigenvalues and eigenvectors) as long as the low-energy field theory is Lorentz invariant. 

\subsection[Topological matter and the Li-Haldane conjecture]{Entanglement Hamiltonian of topological matter and the Li-Haldane conjecture}

As mentioned in the introduction, the structure of the EH for massive phases in condensed matter systems was motivated by the investigation of the connection between entanglement and topological phases. This connection was first formulated in Ref. \cite{Li2008} in the context of fractional quantum Hall states. In particular, it was conjectured that the low-lying entanglement spectrum of a connected partition features the same properties of the edge modes of the theory correspondent to the given state. This observation was backed up by numerical simulations, and has later been verified for a series of other situations in both 1D and 2D systems. For a review, see Ref.~\cite{Laflorencie2016}.

As far as the entanglement Hamiltonian is concerned, the Li-Haldane conjecture strongly suggests that the corresponding EH is dominated, at low-energies, by the edge Hamiltonian. This result is very much consistent with what one would expect from the BW theorem: indeed, this argument was proposed in Ref.~\cite{2012Senthil} as a proof of the connection between edge mode spectra, and entanglement spectra in half-infinite systems. In 1D systems, a further insight can be gathered by directly studying the perturbed CFT: in the case of a relevant perturbation, it has been shown that the corresponding EH is given in terms of the chiral Virasono operator - that, again, is consistent with BW theorem~\cite{cho2017} (this fact can be interpreted similarly to the case of corner transfer matrix in the context of integrable models). This fact has been confirmed on the lattice on several models, including the Haldane and the Su-Schrieffer-Heegeer chain~\cite{cho2017,Giudici2018,TMS2019}.

\subsection[Two-dimensional phases and critical points]{Entanglement Hamiltonian of two-dimensional phases and critical points}

The Li-Haldane conjecture motivated the study of the structure of EH well beyond the one-dimensional case - in particular, for 2D systems. In those settings, considerably less is known, since there are no available explicit expressions in general field theory for finite volumes. 

Direct tests of the accuracy of the lattice BW results in any dimension larger than one are extremely challenging for interacting systems, as they often require the exact diagonalization of the EH. However, it is possible to perform indirect tests based on correlation functions, by performing quantum Monte Carlo (or, in principle, tensor network) simulations of the corresponding model. The first examples in this direction have been the 2D Heisenberg model and Heisenberg bilayer~\cite{Giudici2018,TMS2020}. In both cases, it was found that correlation functions obtained utilizing a 'thermal' sampling of the lattice BW EH asymptotically converge to the value obtained by sampling the original Hamiltonian, substantiating the validity of the approach (even if a finite size scaling theory on how this works is presently lacking).

Based on these insights, the lattice BW EH was utilized to predict properties that are inaccessible otherwise. One example is the von Neumann entropy, that cannot be computed in conventional MC simulations, and whose computation for 2D models with DMRG methods can only be carried out for cylinders of limited width. The advantage of the EH based approach is that the computation of the von Neumann entropy at $T=0$ is cast as a computation of the entropy at finite temperature, but with respect to the EH. This can be conveniently carried out utilizing metadynamics - for instance, with a Wang-Landau algorithm. 

\begin{figure}[t]
\center
\includegraphics[width=0.85\textwidth]{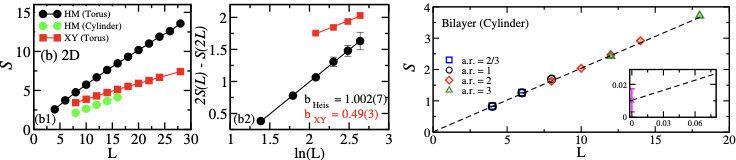}
\caption{Entanglement entropies of 2D lattice spin models obtained via metadynamics Monte Carlo sampling of the lattice Bisognano-Wichmann entanglement Hamiltonian~\cite{TMS2020}. The left and right panel show the scaling of bipartite entropy versus size in Heisenberg, XY, and Heisenberg-bilayer models. The central panel singles out logarithmic corrections, whose coefficient is proportional to the number of Goldstone modes in a phase that features spontaneous symmetry breaking of a continuous symmetry~\cite{Metlitski2011}. The panels of this figure were published in Ref.~\cite{TMS2020}.}
\label{fig:2DEH}
\end{figure}

A sample of the results obtained via the aforementioned method is displayed in Fig.~\ref{fig:2DEH}. The first and third panel show the entanglement entropy of a partition of size $L/2\times L$ in a torus of size $L\times 2L$ for the Heisenberg model (HM), XY model (XY), and bilayer Heisenberg model at the critical point (panel on the right). In the first two cases, the effective field theory describes the spontaneous symmetry breaking of a global $SO(3)$ and $U(1)$ symmetry respectively, while the third case is captured at low-energies by a $O(3)$ sigma model. The results prove that, for these models, the dominant contribution to entropy is linear. 

A finer analysis can be performed for the case of spontaneous symmetry breaking. For those ground states, it was predicted~\cite{Metlitski2011} that the scaling of the entropy shall follow:
\begin{equation}
    S(L) = aL + b \ln(L) + d, \qquad b = n_b/2
\end{equation}
with $a,d$ non universal constants, and $n_b$  the number of Goldstone bosons describing the symmetry-breaking pattern. In the central panel of Fig.~\ref{fig:2DEH}, we show a finite-entropy difference that allows to isolate the logarithmic term by canceling out the area law ones. A clear linear scaling (in $\ln L$) is observed, and the corresponding linear fit returns values of $n_b$ in agreement with field theory predictions. We note that these results are well beyond what can be computed with state of the art tensor network methods, since the values of the corresponding entanglement entropies would require - for the case of matrix-product-state wave functions - prohibitive bond dimensions. This demonstrates how leveraging on the BW EH enables computational methods that allows to make predictions in otherwise inaccessible regimes.

\subsection{Entanglement Hamiltonians out of equilibrium}

So far, we have solely treated equilibrium systems. Unfortunately, extensions of the BW theorem for real time are not known. The only exception is CFT, where it is possible, for certain quench protocols, to identify the EH correspondent to a finite partition~\cite{CardyTonni16,Wen_2018}. Typically, such EH depend on the full stress energy tensor (not only on its energy density component, $T_{00}$), so their direct application to lattice models requires more caution. Still, some results exist for the harmonic and free-fermion chains \cite{DGAT19}, and we will see in the next section one working example in the context of quantum spin chains.

\section{Entanglement Hamiltonians in experiments}
\label{sec:Exp}

We have introduced in the previous sections closed form formulas for the entanglement Hamiltonian of lattice models.
The fact that the entanglement Hamiltonian can be expressed in a form that is reminiscent of the original Hamiltonian allows us to devise methods to probe the entanglement Hamiltonian experimentally.
Below we review recent theoretical and experimental works that showed how to make use of the concept of entanglement Hamiltonians in order to experimentally extract the entanglement properties of a many-body system.

We organize this section in three parts:
In Sec.~\ref{sec:Exp_Imp}, we show how one can use the \emph{physical implementation} of the entanglement Hamiltonian in a experiment as a scalable method to measure the entanglement spectrum.
We then show in Sec.~\ref{sec:Exp_Meas} how one can efficiently \emph{measure} the entanglement Hamiltonian based on local ansatze that are inspired from the BW theorem.
These two approaches are finally combined in Sec.~\ref{sec:Exp_Imp_Meas}.
In addition, and in order to help the reader in comparing these methods, we provide in Table.~\ref{table:Exp} a summary for each protocol of the different experimental requirements, and of quantities that can be measured. 

\subsection{Implementation of entanglement Hamiltonians}
\label{sec:Exp_Imp}
The first motivation behind the approach of quantum simulation of entanglement Hamiltonians presented in Ref.~\cite{Dalmonte2018} is the measurement of entanglement spectra (ES).
The question of the experimental access to the ES is a long-standing challenge for quantum simulation. Measuring the ES allows us, for instance, to understand the structure of symmetry-protected-topological phases~\cite{Pollmann2010}, or more generally to test the Li-Haldane conjecture~\cite{Li2008,Regnault2015}.
The measurement of the ES can also be used to extract the entanglement entropies that quantify the presence of entanglement in correlated quantum systems~\cite{Eisert2010}.

In the standard scenario of quantum simulation~\cite{Georgescu2014},
a quantum state of interest $\ket{\psi}$ is  realized based on the physical implementation of the Hamiltonian $H$ of a lattice model.
This implementation can be based on ultracold atoms, ions, or superconducting circuits realizing a quantum computer, etc.
For instance, we can be interested in the entanglement properties of the ground state $\ket{\psi_0}$ of $H$, and we would like to probe the reduced density matrix $\rho_A=\mathrm{Tr}_{S-A}(\ket{\psi_0}\bra{\psi_0})$ of the subsystem $A$. Having access to the ES, the eigenvalues of $\rho_A$,  is notoriously difficult in quantum simulation or quantum computing. One either needs to measure the full density matrix $\rho_A$ via quantum state tomography, which requires exponentially many measurements~\cite{Haah2017}. Alternatively, one can use interferometric methods that have the drawback of requiring a large number of multiple copies of the system~\cite{Pichler2016}. We show for illustration in Fig.~\ref{fig:Exp_Imp}a)-b) a measurement of the ES performed in a quantum computer~\cite{Choo2018}, and obtained via quantum state tomography.
In Fig.~\ref{fig:Exp_Imp}a), the quantum circuit to create the groundstate of a symmetry protected topological phase with $L=8$ is represented. Quantum state tomography is then realized on a reduced density matrix of $\ell=4$ qubits, which gives the ES shown in Fig.~\ref{fig:Exp_Imp}b). 

\begin{table*}
\begin{tabular}{| c | c|c | r |}
  \hline		
  
  Method & States implemented & Types of measurement & Measured quantities \\
  \hline \hline
  Quantum simulation of EHs	~\cite{Dalmonte2018} & The eigenstates of $\mathcal{H}$
  & Spectroscopy  &
   Entanglement spectrum \\
  \hline
  EH tomography~\cite{Kokail2021} & $\ket{\psi}$ with reduced states  $\rho_A$&
  Tomography on $\rho_A$
  & Entanglement Hamiltonian \\
   &&& Entanglement spectrum \\
  \hline
  Variational estimation of the EH~\cite{Kokail2021quantum} & $\ket{\psi}$, with $\rho_A$ time evolved &
Time-evolved observables &Entanglement Hamiltonian  \\
    & with estimations of $\mathcal{H}$ && Entanglement spectrum \\
  \hline  
\end{tabular}
\caption{Summary of methods for probing entanglement Hamiltonians. Ref.~\cite{Kokail2021quantum} combines Ref.~\cite{Dalmonte2018} and Ref.~\cite{Kokail2021} by considering a variational learning procedure of the EH, where estimations of the EH that are iteratively updated are physically applied on the state $\rho_A$.}
\label{table:Exp}
\end{table*}

Beyond small qubit sizes, full state tomography is no longer an option due to the exponential cost in terms of measurements.
In order to access easily ES for large systems, it was proposed in Ref.~\cite{Dalmonte2018} to physically implement the entanglement Hamiltonian $\mathcal{H}$ in an experiment.
Instead of preparing the ground state $\ket{\psi_0}$ via the implementation of $H$, as in the `traditional' scenario of quantum simulation, one prepares the eigenstates $\ket{\tilde{\psi}_{\alpha}}$ of the entanglement Hamiltonian $\mathcal{H}$.
As shown below, such quantum simulation of the entanglement Hamiltonian can be used in particular to extract, in a scalable way, the entanglement spectrum using, well established, spectroscopy techniques.

\subsubsection{Implementing the entanglement Hamiltonian}
The idea behind Ref.~\cite{Dalmonte2018} is that the BW theorem provides an experimentally-friendly method to implement entanglement Hamiltonians.
In particular, the lattice version of the BW theorem Eq.~\eqref{BW-EH} gives an expression of the entanglement Hamiltonian $\mathcal{H}$  in terms of  particles operators (spins, bosons/fermions), which we can relate in the context of an experiment to the particles of the system (cold atoms, etc). 
Also, the locality of $\mathcal{H}$ copes well with the local character of the interactions that can be implemented between atoms, ions, etc.
Finally, as these interactions are often mediated using tunable controls, such as external lasers, we can easily engineer inhomogeneous terms following the prescription of the BW theorem. This is illustrated in Fig.~\ref{fig:Exp_Imp}c)-d) for Rydberg and trapped ions quantum technologies implementing the entanglement Hamiltonian of quantum Ising models.
Using cold atoms placed in optical lattices, the EH can be engineered via laser-assisted tunneling~\cite{Dalmonte2018}, or other types of inhomogeneous optical forces~\cite{Barfknecht2021}.

\subsubsection{Entanglement Hamiltonian spectroscopy}
\begin{figure}
\centering
\includegraphics[width=0.8\columnwidth]{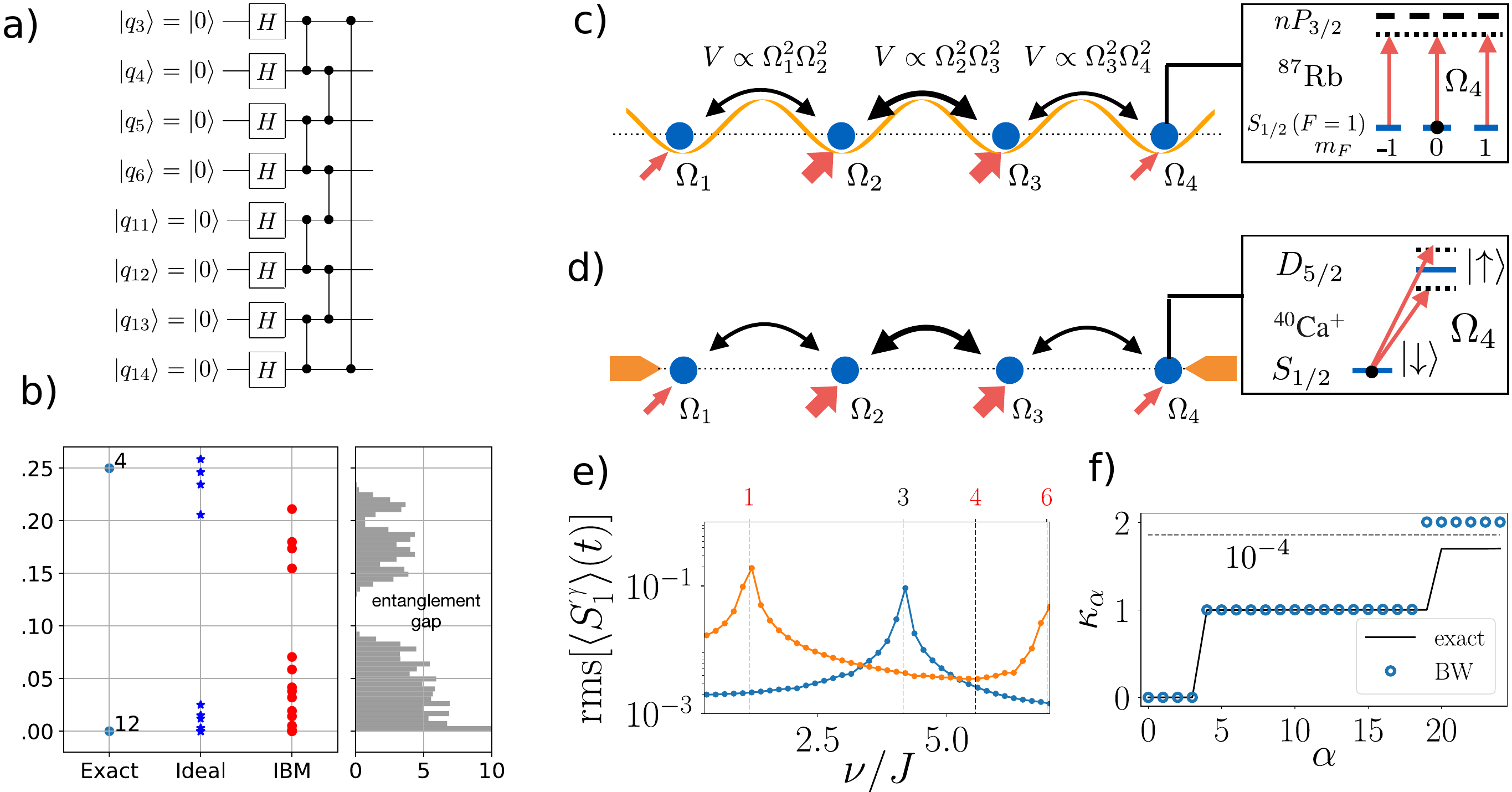}
\caption{
(a-b) Extraction of ES from quantum state tomography in a quantum computer~\cite{Choo2018}.
a) Quantum circuit to prepare a symmetry protected topological groundstate with $8$ qubits. The symbol $H$ indicates the Hadamard gate, which is a single qubit gate. The double dot symbol refers to the two qubit controlled-$Z$ gate.
b) Corresponding ES extracted via tomography, showing the $4$-fold ES degeneracy of the Haldane phase.
(c-f) Quantum simulation and spectroscopy of entanglement Hamiltonians~\cite{Dalmonte2018}.
c)-d) Implementations of entanglement Hamiltonians using interactions mediated with inhomogeneous laser beams in Rydberg and trapped ions platforms.
e) Entanglement spectroscopy of the Haldane phase. The positions of the peaks signal the eigenvalues of the entanglement Hamiltonian. Here, we consider $\ell=8$ and $L=100$.
f) Normalized entanglement spectra obtained for a large sub-system of $\ell=40$ lattice sites.
Panels (a)-(b) were published in Ref.~\cite{Choo2018}.
Panels (c)-(f) were published in Ref.~\cite{Dalmonte2018}.}
\label{fig:Exp_Imp}
\end{figure}

Once the EH is implemented in an experiment, one can explore  entanglement properties, in a scalable way, and using the standard tools of quantum simulation.
In particular, in order to reveal the ES experimentally, one can first prepare the ground state $\ket{\tilde \psi_0}$ with energy $\epsilon_0$ of $\mathcal{H}$.
The spectroscopy of $\mathcal{H}$ is then realized by applying a weak perturbation generated by a time-dependent Hamiltonian $h(t)=h \sin(\nu t)$.
When the frequency of the perturbation approaches an entanglement `transition', $\nu\approx \epsilon_{\alpha}-\epsilon_0$, the corresponding eigenstate $\ket{\tilde \psi_{\alpha}}$ is resonantly excited, and we observe a response of the system that can be detected by monitoring the dynamics of certain observables. In Fig.~\ref{fig:Exp_Imp}e), this is illustrated for the case of the entanglement spectroscopy of the spin-$1$ Haldane phase~\cite{Pollmann2010}.
Here, the value of the first entanglement transition vanishes in the thermodynamic limit, corresponding to the presence of topological degeneracies in the entanglement spectrum, see Fig.~\ref{fig:Exp_Imp}f) for a subsystem of $\ell=40$ sites. This means that, by monitoring the closure of the entanglement gap in such an experiment, we can reveal the topological nature of the Haldane phase.
Utilizing similar ideas, it has been proposed in Ref.~\cite{TMS2020} that also moments of the ES distribution - such as the von Neumann entropy - can be measured.

\subsection{Measurements of entanglement Hamiltonians}
\label{sec:Exp_Meas}

As described above, under of the assumptions of the BW theorem, we can reveal entanglement properties by  physically implementing the EH $\mathcal{H}$.
In certain situations however, e.g., away from equilibrium, the BW theorem does not strictly apply. 
Therefore we may wonder whether an experiment is able to \emph{measure} the entanglement Hamiltonian, and in particular check the validity of the BW theorem.
Here, in contrast to Sec.~\ref{sec:Exp_Imp}, we have in mind again the `traditional' scenario of quantum simulation where an arbitrary state $\ket{\psi}$ with reduced density matrices $\rho_A$ is physically realized, c.f., Table.~\ref{table:Exp}. 
The goal is to measure the corresponding entanglement Hamiltonian.

As the EH $\mathcal{H}$ is the matrix log of the density matrix $\rho_A$, the measurement of $\mathcal{H}$ is formally equivalent to quantum state tomography, and thus require in principle exponentially many measurements~\cite{Haah2017}.
In Ref.~\cite{Kokail2021}, it was however shown  that one can use the concept of entanglement Hamiltonians to perform quantum state tomography more efficiently.

\begin{figure}
 \centering
\includegraphics[width=0.85\columnwidth]{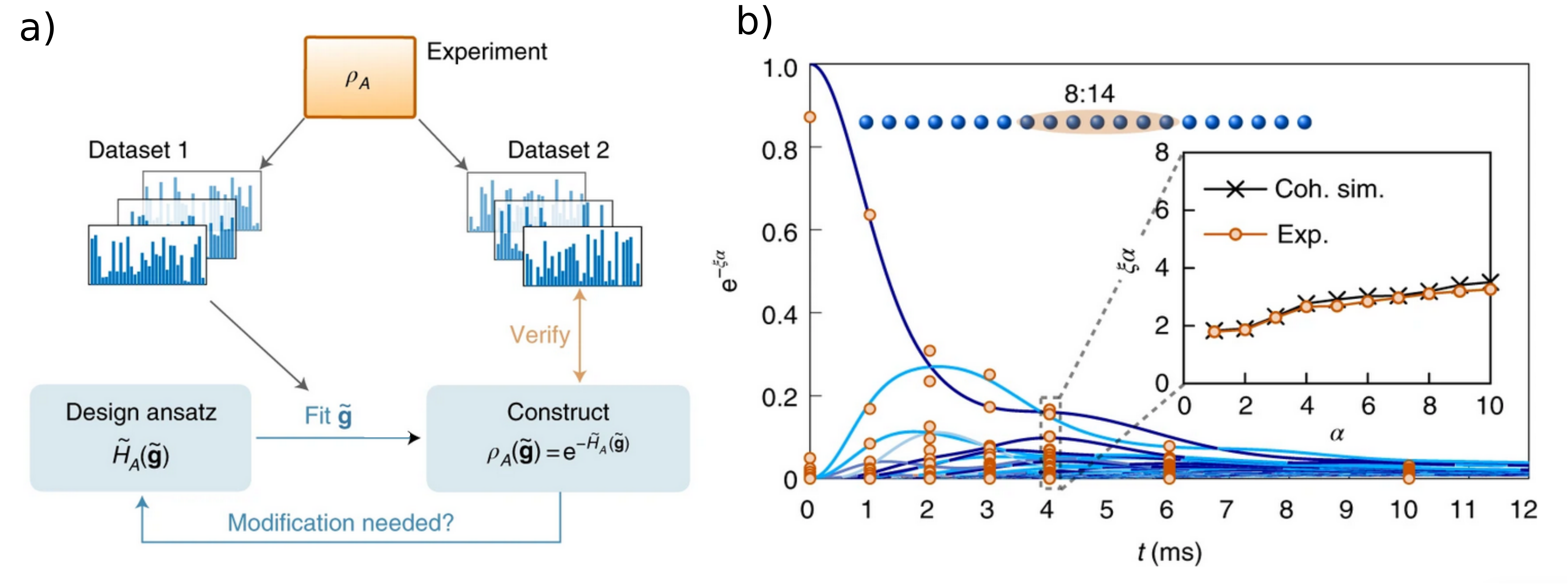}
\caption{Measurement of entanglement Hamiltonians.
a) Protocol for EHT~\cite{Kokail2021} based on fitting estimations of the density matrix with an ansatz based on a local EH.
b) Experimental reconstruction from EHT of the ES from the randomized measurements data of Ref.~\cite{Brydges2019}.
Figure published in Ref.~\cite{Kokail2021}. }
\label{fig:Exp_Meas}
\end{figure}

The entanglement Hamiltonian tomography (EHT) protocol is shown in Fig.~\ref{fig:Exp_Meas}a).
The system is subject to various measurements, which provide an estimation of the density matrix $\rho_A$. 
This estimation is typically inaccurate because made from too few measurements compared to the requirements of tomography. 
However, this `poor' density matrix estimation is then fitted to an ansatz $\rho_A\propto e^{-\mathcal{H}}$ that is based on a local EH $\mathcal{H}$, i.e., inspired from the result of the BW theorem. As this ansatz is made of a polynomial number of fitting terms in system size, very few measurements are indeed necessary to obtain a faithful reconstruction of the density matrix.

Interestingly, the process of using experimental measurements to reconstruct the entanglement Hamiltonian can be realized in the framework of randomized measurements that provide estimations of density matrices~\cite{Ohliger2013,Elben2019} (among other quantities, see below).
In Ref.~\cite{Kokail2021}, this was used to measure entanglement Hamiltonians and spectra, based on  existing randomized measurements experimental data from Ref.~\cite{Brydges2019}. The corresponding experimental reconstructing of the ES is shown in  Fig.~\ref{fig:Exp_Meas}b). 
Here the system under study was a trapped ion spin chain that was evolved via the long-range $XY$ Hamiltonian.

Remarkably, the efficiency of entanglement Hamiltonian tomography only relies on the fact that the EH is local. 
In this case, the density matrix can be interpreted as a Gibbs state of a local Hamiltonian (the EH), which, as shown rigorously in Ref~\cite{Anshu2021}, can be indeed `learned' from few measurements.
Importantly, the method presented in Ref.~\cite{Kokail2021} can be used to \emph{prove} the locality of the EH, and in particular for ground states to verify the lattice version of the BW theorem.
Using a randomized measurement protocol for fidelity measurements~\cite{Elben2020} as sub-routine, one can check that the EH is indeed faithfully estimated.
If necessary, additional terms can also be included to `enrich' the EH and to provide a better fit to the experimental data. 

\subsection[Combined implementations and measurements]{Combined implementations and measurements of entanglement Hamiltonians}\label{sec:Exp_Imp_Meas}

\begin{figure}
 \centering
\includegraphics[width=0.85\columnwidth]{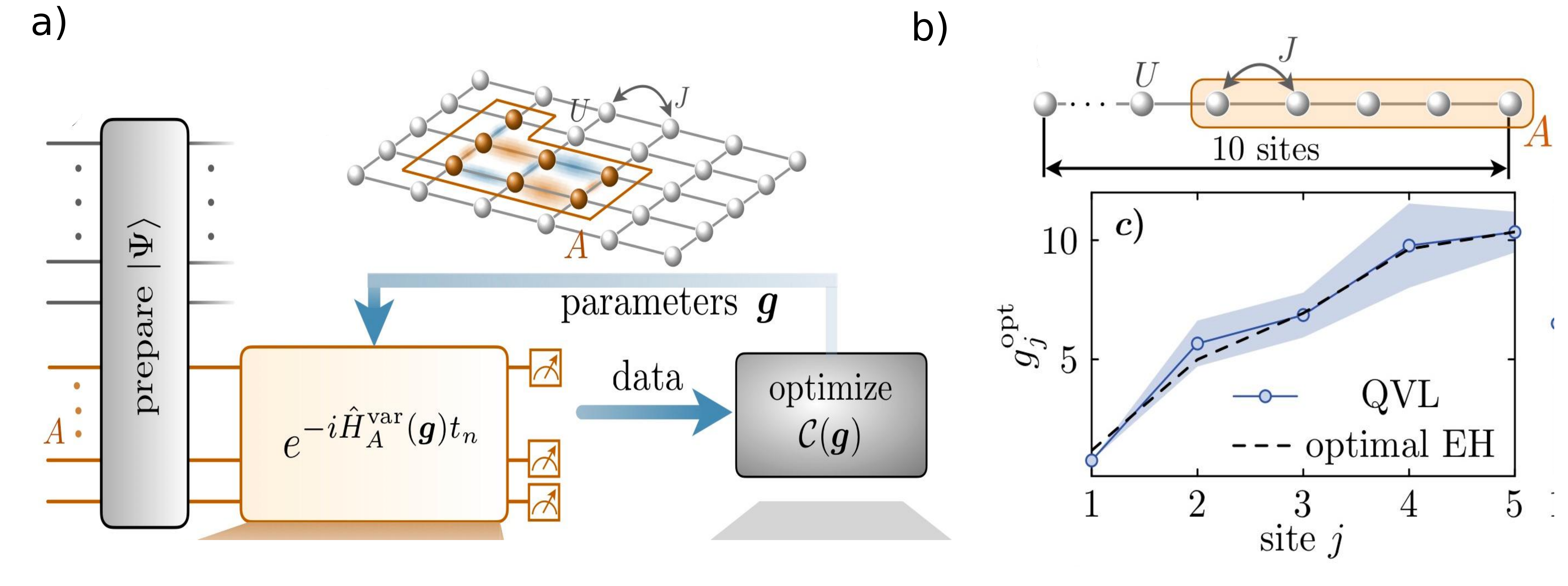}
\caption{Measurement of entanglement Hamiltonians.
a) Quantum variational learning of the entanglement Hamiltonian~\cite{Kokail2021quantum}. This is based on a minimization procedure using as input a measurement of certain observables, and as outputs the variables that parametrize the EH.
b) Numerical illustration of the protocol, where the local terms of the EH are reconstructed at the end of the minimization procedure.
Figure published in Ref.~\cite{Kokail2021quantum}.}
\label{fig:Exp_Variational}
\end{figure}

The protocol~\cite{Dalmonte2018} physically implements the EH $\mathcal{H}$ according to the prescription of the BW theorem. Instead , the protocol Ref.~\cite{Kokail2021} measures the EH via a tomography of the density matrix $\rho_A$, and can in particular signal deviations from the BW theorem.
Recently, the advantages of these two approaches have been combined in a new protocol presented in Ref.~\cite{Kokail2021quantum}.
In this approach, one iteratively measures the EH, \emph{and} implements it on the quantum device in order to perform the entanglement spectroscopy. 
As shown in Fig.~\ref{fig:Exp_Variational}a), the idea is to run an  iterative algorithm, where at each step $n$: (i) we  first 
realize a quantum state $\ket{\psi}$, 
(ii) we  apply on a subsystem $\rho_A$ time evolution with an estimation \mbox{$\mathcal{H}[n]=\sum_i \mathcal{H}_i[n]$} of the EH, which is parametrized with a polynomial number of local terms $\{\mathcal{H}_i[n]\}$. 
Note that such parametrization also occurs in the context of Hamiltonian learning~\cite{Bairey2019,Evans2019,Qi2019,Li2020}.
(iii) we measure the response of the system,  and assess how the estimation $\mathcal{H}[n]$ of the unknown EH can be improved, see below. 
The sequence (i-ii-iii) is then repeated, until the algorithm converges. At this point, we obtain a  faithful estimation of the EH, which can be then used for entanglement spectroscopy.

The procedure to update the estimation of the EH from a measurement, which is the crucial step (iii) of the protocol, builds from the following observation: 
If the initial state $\rho_A$ has been time evolved from $\mathcal{H}$, the final state  must remain unchanged, i.e  $\rho_A(t)=e^{-i\mathcal{H} t} \rho_A e^{i\mathcal{H} t}=\rho_A$, because $\rho_A$ and $\mathcal{H}$ commute. Therefore, we can translate the problem of finding $\mathcal{H}$ into a minimization problem. The cost function to be minimized is defined as $\mathcal{C}_n(t)=|\langle O_n(t) \rangle-\langle O_n(0) \rangle|$,  where \mbox{$\langle O_n(t) \rangle=\mathrm{Tr}(O e^{-i\mathcal{H}[n] t} \rho_A e^{-i\mathcal{H}[n] t})$} and $O$ is a well chosen observable. The variables to be adjusted are the local terms $\{\mathcal{H}_j\}$.
Such minimization procedure can be executed on a classical computer, using as input the experimental measurements of the cost function $\mathcal{C}_{m=0,\dots,n}(t)$, and as outputs the variables $\{\mathcal{H}_i[n+1]\}$ that parametrize the new EH candidate to be tested on the experiment. A numerical example of reconstruction of local terms parametrizing the EH is shown in Fig.~\ref{fig:Exp_Variational}b).

\subsection{Summary of the experimental section and comparisons with other methods}

We have presented experimental protocols that take advantage of the concept of entanglement Hamiltonians to probe
entanglement properties of many-body systems. 
These probing methods are scalable in the sense that one requires a number of measurements that scales polynomially in system size. 
This is much less than the number of measurements required for state-tomography, which typically scale as $r^2D$ for density matrices $\rho_A$ of rank $r$ and dimension $D$~\cite{Haah2017}.
As a direct consequence, while the ES was previously measured based on tomography~\cite{Choo2018} on moderate system sizes of $4$ qubits, the protocols presented above show how to measure the ES in much larger systems.

Having access to the ES and EH give also access to any entanglement property, such as entanglement entropies.
Compared to alternative methods based on 
 implementing multiple copies to measure R\'enyi entropies~\cite{Alves2004,Daley2012,Islam2015,Kaufman2016}, 
negativities~\cite{Gray2017}, (and also the ES~\cite{Pichler2016}), the protocols presented above involve single instances of the quantum states and appear thus as more experimentally friendly.
Finally, protocols based on measuring statistical correlations between randomized measurements have been also devised to measure  entanglement entropies~\cite{vanEnk2012,Elben2018,Brydges2019,Huang2020,Rath2021importance,Vitale2021}, and entanglement negativities~\cite{Zhou2020,Elben2020mixed}.
These protocols have the advantage of not making assumptions on the state, and of not requiring multiple copies.
However, they require, in experiments with $N$ qubits, an exponential number of measurements 
$2^{\alpha N}$, with a modest exponent $\alpha \le 1$ compared to tomography. 
As described above, randomized measurement protocols can also be combined with EH tomography~\cite{Kokail2021}.

\section{Conclusions and outlook}

The characterization of quantum correlations in quantum many-body systems is of paramount importance to deepen our understanding of physical phenomena, with direct applications to quantum technologies. In this review, we have summarized how the entanglement Hamiltonian constitutes an extremely powerful tool to carry out such characterization. Our presentation has followed the three parallel lines that have characterized its study: quantum field theory, integrable systems, and topological matter. Those three lines - that have had very little overlaps for almost thirty years - have been mutually intersecting over the last decade, providing deep insights on the structure of equilibrium state of matter. We have tried to emphasize as much as possible such intersections, and in particular, the pivotal role played by exact results in both axiomatic quantum field theory and integrable systems along these developments. 

The overarching message that this review summarizes is that, for most physical states of interest to many-body theory, the entanglement structure of bipartitions simplifies dramatically, and is in fact captured by {\it local} (inhomogeneous) operators, featuring only few-body terms, similarly to conventional Hamiltonian dynamics. Such structure gives direct access to both a physical interpretation of entanglement (via, e.g., Unruh effect), and paves the way to powerful applications in quantum information processing, including methods to measure entanglement spectra and even perform quantum state tomography based on the simplified structure of the entanglement Hamiltonian - methods that have already been  demonstrated on experimental data. 

There are a number of open questions in the field. A first set of questions is related to the structure of the entanglement Hamiltonian in different setups from the ones described here. One important example is, what is the structure of entanglement Hamiltonians out of equilibrium: here, the present understanding is limited to specific quench protocols in conformal field theory, and a few numerical examples. Deeper insights on the entanglement structure after quench may help shedding light on the origin of the so-called entanglement barrier, and on the evolution of operator entanglement along thermalization, just to name two applications.

Another field where little is known is the case of disconnected partitions: while it is known that entropies of such partitions are more informative that the ones of connected partitions for specific cases \cite{Gliozzi2008,Furukawa2009,CCT09}, very little is known about the corresponding EH, except for free field theories (see Sec. \ref{sec:CFTfurther}). Beyond numerical studies, it would be very intriguing to see whether it is possible to at least partly address such questions in the context of axiomatic quantum field theory: exact results such as the Bisognano-Wichmann theorem have proven pivotal so far, and one expects similar impact on the cases above. Another set of questions concerns applications that leverage on the known (or assumed) structure of entanglement Hamiltonians to empower tools in quantum information processing. While it is challenging to speculate on future applications, the successful example of tomographic reconstruction methods (one if not the the most challenging characterization of quantum states) is definitely a strong basis to build future applications on. 

Until recently, the study of entanglement Hamiltonians had followed separate paths in different branches of mathematics and (theoretical) physics. Now that links between these fields have been established, we expect an even faster and more efficient crossfertilization to take place, aimed at characterizing quantum correlations in many-body systems from an operatorial viewpoint.

\medskip
\textbf{Acknowledgements} \par 

We thank P. Calabrese, A. Elben, R. Fazio, G. Giudici, C. Kokail, T. Mendes-Santos, S. Murciano, I. Peschel, M. A. Rajabpour, B. Sundar, E. Tonni, R. Van Bijnen, V. Vitale, T. Zache, J. Zhang, and P. Zoller for discussions and collaborations over the years on topics related to the present review.
VE acknowledges funding from the Austrian Science Fund (FWF) through Project No. P35434-N.
BV acknowledges funding from the Austrian Science Fund (FWF, P 32597 N), and from the French National Research Agency (ANR-20-CE47-0005, JCJC project QRand).
MD is partly supported by the ERC under grant number 758329 (AGEnTh), by the MIUR Programme FARE (MEPH), and by European Union's Horizon 2020 research and innovation programme under grant agreement No 817482 (Pasquans).

\bibliographystyle{MSP}
\bibliography{mathbib.bib}

\end{document}